\documentclass[prd,preprintnumbers,twocolumn,eqsecnum,floatfix,a4paper,nofootinbib]{revtex4}
\usepackage{color}
\usepackage{calc}
\usepackage{amsmath,amssymb,graphicx}
\usepackage{amssymb,amsmath}
\usepackage{tensor}
\usepackage{bm}
\usepackage{microtype}
\usepackage{booktabs}
\usepackage{times}
\usepackage[varg]{txfonts}
\usepackage[colorlinks, pdfborder={0 0 0}]{hyperref}
\usepackage{subfig}
\usepackage{tabularx}
\usepackage{footmisc}
\definecolor{LinkColor}{rgb}{0.75, 0, 0}
\definecolor{CiteColor}{rgb}{0, 0.5, 0.5}
\definecolor{UrlColor}{rgb}{0, 0, 0.75}
\hypersetup{linkcolor=LinkColor}
\hypersetup{citecolor=CiteColor}
\hypersetup{urlcolor=UrlColor}
\allowdisplaybreaks
\DeclareFontFamily{OT1}{pzc}{}
\DeclareFontShape{OT1}{pzc}{m}{it}{<-> s * [1.10] pzcmi7t}{}
\DeclareMathAlphabet{\mathpzc}{OT1}{pzc}{m}{it}

\newcommand{\hlm}{\mathpzc{h}_{\ell m}}

\newcommand{\Mo}{M_{\odot}}
\newcommand{\FFe}{\mathrm{FF}_\mathrm{eff}}
\newcommand{\FF}{\mathrm{FF}}

\newcommand{\rhoopt}{\rho_\mathrm{opt}}

\newcommand{\TemplateName}{IMRPhenomD}
\newcommand{\chieff}{\chi_\mathrm{eff}}
\newcommand{\mchieff}{\tilde{\chi}_\mathrm{eff}}
\newcommand*{\skymapscale}{0.235}

\newcommand*{\minsnrscale}{0.44}

\begin{document}

\newcommand{\be}{\begin{equation}}
\newcommand{\ee}{\end{equation}}
\newcommand{\ber}{\begin{eqnarray}}
\newcommand{\eer}{\end{eqnarray}}
\def\bea{\begin{eqnarray}}
\def\eea{\end{eqnarray}}
\newcommand{\etal}{\emph{et al}}

\title{Effects of nonquadrupole modes in the detection and parameter estimation of \\black hole binaries with nonprecessing spins}
\author{Vijay~Varma}
\email{vvarma@caltech.edu}
\affiliation{Theoretical Astrophysics, California Institute of Technology, Pasadena, California 91125, USA}
\author{Parameswaran~Ajith}
\affiliation{International Centre for Theoretical Sciences, Tata Institute of Fundamental Research, 
Bangalore 560089, India}
\affiliation{Canadian Institute for Advanced Research, CIFAR Azrieli Global Scholar, 
MaRS Centre, West Tower, 661 University Avenue, Suite 505, Toronto, Ontario M5G 1M1, Canada}

\begin{abstract}
We study the effect of nonquadrupolar modes in the detection and parameter estimation 
of  gravitational waves (GWs) from black hole binaries with nonprecessing spins, 
using Advanced LIGO. We evaluate the loss of the signal-to-noise ratio (SNR) and the systematic 
errors in the estimated parameters when a quadrupole-mode template family is used to detect 
GW signals with all the relevant modes. Target signals including nonquadrupole modes are 
constructed by matching numerical-relativity simulations of nonprecessing black hole binaries 
describing the late inspiral, merger and ringdown with post-Newtonian/effective-one-body 
waveforms describing the early inspiral. We find that neglecting nonquadrupole modes will, 
in general, cause unacceptable loss in the detection rate and unacceptably large systematic 
errors in the estimated parameters, for the case of massive binaries with large mass ratios. 
For a given mass ratio, neglecting subdominant modes will result in a larger loss in the 
detection rate for binaries with aligned spins. For binaries with antialigned spins, 
quadrupole-mode templates are more effectual in detection, at the cost of introducing a 
larger systematic bias in the parameter estimation. We provide a summary of the regions 
in the parameter space where  neglecting nonquadrupole modes will cause an unacceptable loss 
of detection rates and unacceptably large systematic biases in the estimated parameters.  
\end{abstract}
\preprint{LIGO-P1600332-v2}
\maketitle
\section{Introduction and Summary}
\label{sec:intro}

We are firmly in the era of gravitational wave (GW) astronomy, with LIGO having made two confident 
detections of binary black holes~\cite{LSC_2016firstdetection, LSC_2016seconddetection} and many more expected in 
upcoming observing runs~\cite{LSC_2016rates, LSC_2016O1results}. Indeed, these first observations 
have already given us a glimpse of the unique capabilities of GW astronomy. Apart from providing 
the first direct evidence of the existence of GWs, these observations confirmed the existence of 
stellar mass black holes that are much more massive than commonly thought by 
astronomers~\cite{LSC_2016paramest, LSC_2016astroph}. They also provided the first evidence of black hole 
binaries that inspiral under GW emission and merge within the age of the Universe. These observations 
also enabled us to perform the first tests of GR in the highly relativistic and nonlinear regime of 
gravity -- a regime inaccessible by other astronomical observations and laboratory tests~\cite{LSC_2016grtests}. 

The first LIGO event, termed GW150914, was produced by the merger of two massive black holes. The 
resultant signal in the detectors contained imprints of the late inspiral and merger of the 
two holes and the subsequent ringdown of the remnant black hole. The signal was 
first detected by two low-latency searches for generic transient signals 
that are coherent in multiple detectors~\cite{Klimenko:2015ypf,Lynch_2015BurstSearch,
0264-9381-25-11-114029,LSC_2016burstsearch}. 
The signal was later confirmed with higher confidence by matched filter-based 
searches that use relativistic models of expected signals from coalescing compact 
binaries~\cite{LSC_2016cbcsearch, Canton:2014ena, Usman:2015kfa, Cannon:2011vi}. The second signal was produced by the coalescence of two 
less massive black holes, and the resultant signal in the detector predominantly consisted of the 
long inspiral. Hence matched filter-based searches were essential for its detection~\cite{LSC_2016seconddetection}. 

Matched-filtering is the most sensitive search method for extracting signals of known signal shape
from noisy data, such as the GW signals from the coalescence (inspiral, merger and ringdown) of 
binary black holes. The source parameters are then extracted by comparing the data against theoretical 
templates by means of Bayesian inference~\cite{LSC_2016paramest, Veitch:2014wba}. Our ability to optimally detect the signal 
using matched-filtering and to estimate the source parameters using Bayesian inference 
depends crucially on how faithfully the theoretical templates model the signal present in the data. 
If the template is a poor representation of the true signal, this can reduce the matched-filtering 
signal-to-noise ratio (SNR), potentially causing nondetection and/or causing unacceptable systematic 
biases in the estimated parameters. Good waveform templates should be not only \emph{effectual} in the 
detection  (small loss in the SNR) but also \emph{faithful} in parameter estimation (small systematic 
biases)~\cite{DIS98}.

%
\begin{figure*}[htb]
\begin{center}
    \subfloat[\label{subfig:summary_det} For detection]
    {\includegraphics[scale=0.5]{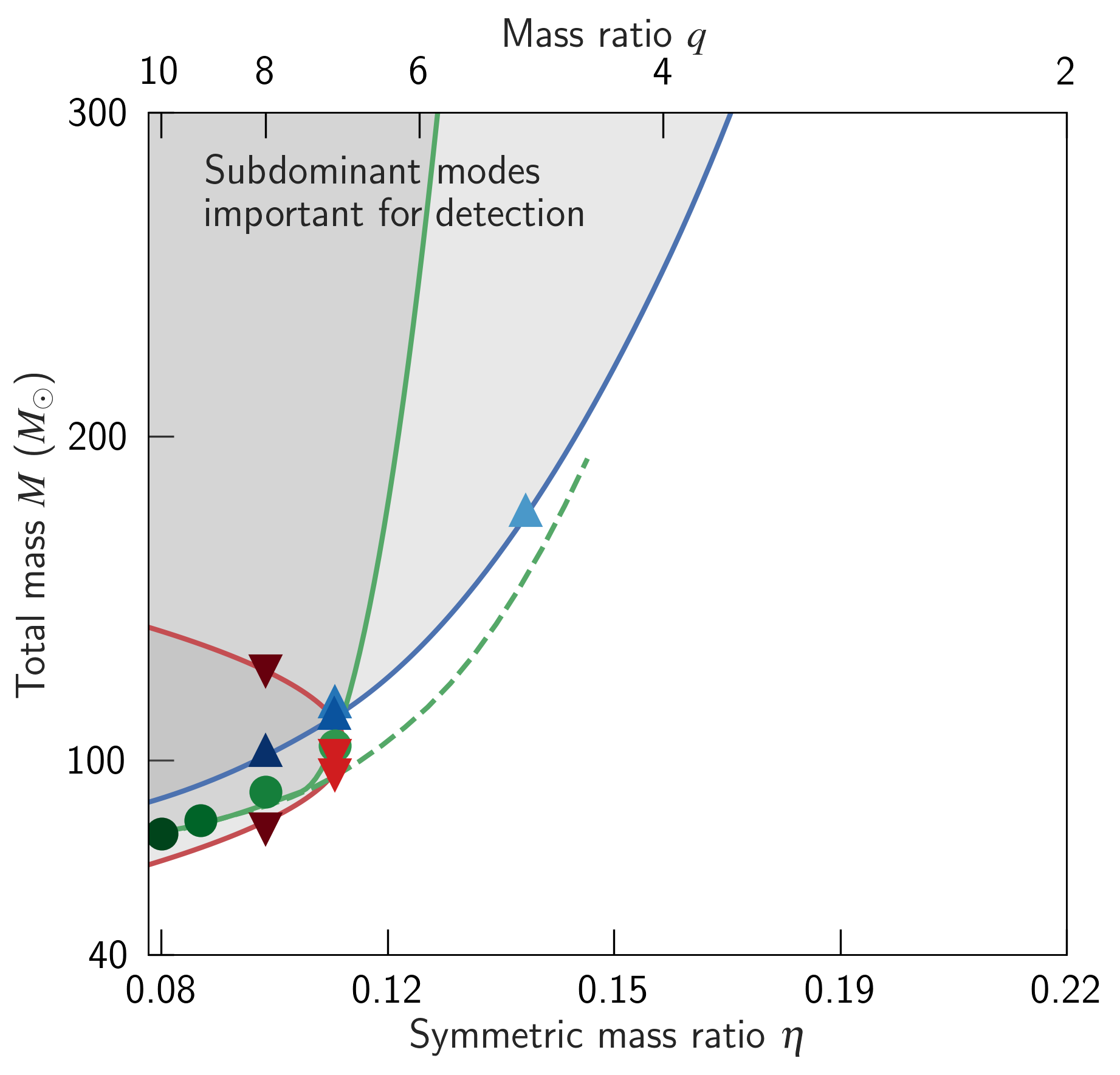}}
    \subfloat[\label{subfig:summary_param} For parameter estimation]
    {\includegraphics[scale=0.5]{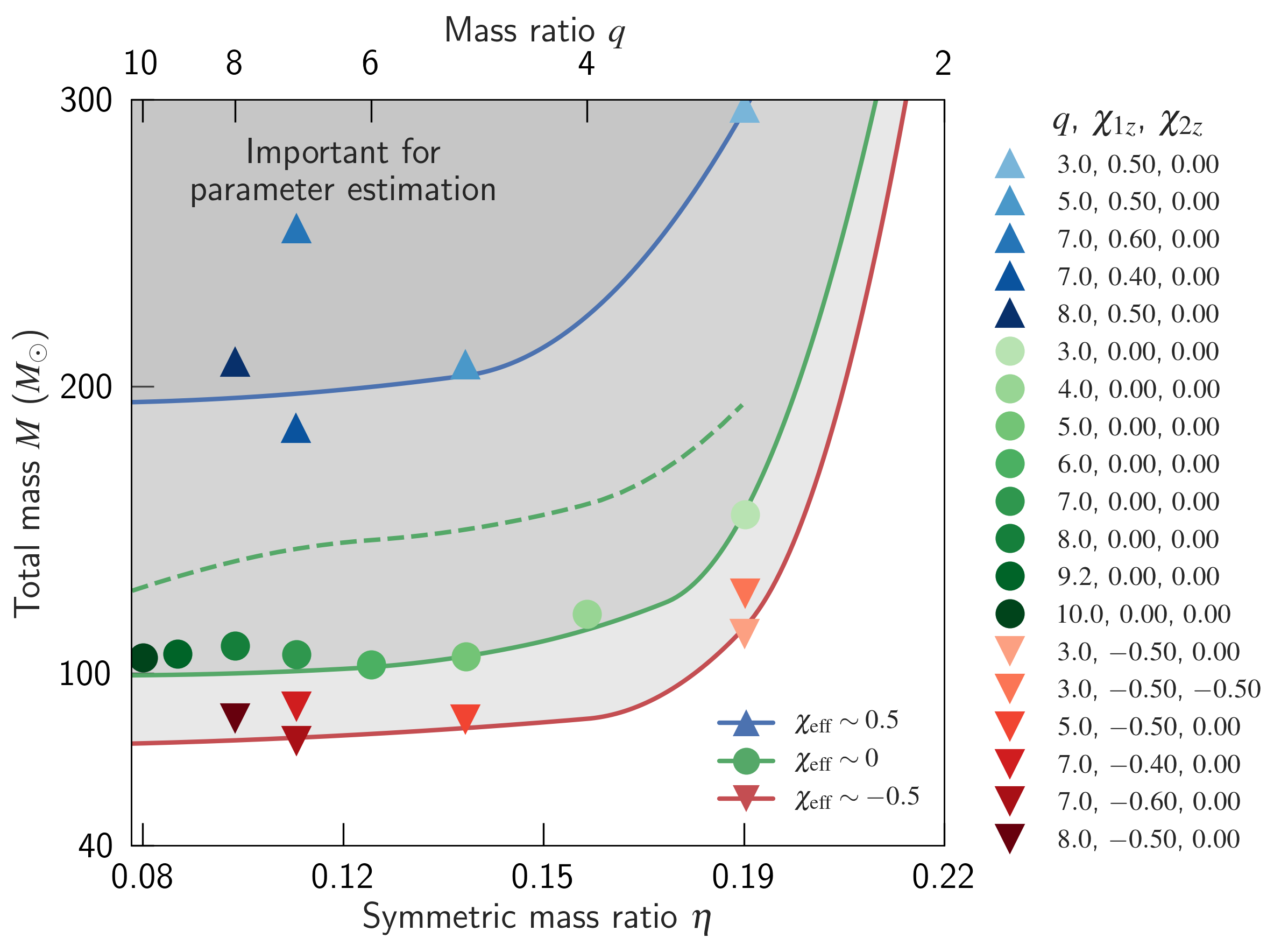}}
\caption{These plots summarize the region in the parameter space of nonprecessing black-hole binaries where contributions from subdominant modes are important for detection (left) and parameter estimation (right).  In the left panel, the shaded areas show the regions in the parameter space where the loss of detection volume (for a fixed SNR threshold) due to neglecting subdominant modes is larger than $10\%$. In the right panel, shaded areas show the regions in the parameter space where the systematic errors in any of the estimated parameters [total mass $M := m_1 + m_2$, symmetric mass ratio $\eta := m_1 m_2/M^2$ and effective spin parameter $\chieff := (m_1 \chi_1 + m_2 \chi_2/M$)] are larger than the expected statistical errors for a sky and orientation-averaged SNR of 8 (corresponding to an optimal orientation SNR $\simeq 20$). In each plot the three solid curves correspond to different effective spin values: blue for $\chieff \sim 0.5$, green for $\chieff \sim 0$ and red for $\chieff \sim -0.5$. The left panel was made by computing the fitting factors of dominant-mode templates including nonprecessing spins with hybrid waveforms including all the relevant modes, and the right panel was made making use of averaged systematic biases. The markers (triangles pointing up/down denoting binaries with aligned/antialigned spins and circles denoting nonspinning binaries) indicate the data points that are used to construct the shaded regions and curves. The legend shows the mass ratios and spins of the target signals featured in these plots. See Sec.~\ref{sec:intro} for a summary and Sec.~\ref{sec:results} for a detailed discussion. For comparison, the dashed green lines show the same results for nonspinning binaries using a nonspinning template family from our previous work~\cite{Varma:2014hm}.}
\label{fig:summary_fig}
\end{center}
\end{figure*}
%

Matched filter-based searches for GWs performed to date, including the ones that resulted in detections, 
have employed templates that model only the leading (quadrupole, or $\ell=2$, $m=\pm2$) spherical harmonic modes of the GWs
radiated from the binary. The parameter estimation exercise also has largely employed quadrupole mode
templates (with the notable exception of one that directly employed numerical-relativity (NR) waveforms~\cite{Abbott:2016apu}). 
This choice is partly dictated by the unavailability of fast-to-evaluate, semianalytical waveform templates describing 
the inspiral, merger and ringdown of binary black holes that model the subdominant (nonquadrupole) modes 
over a sufficiently wide region in the parameter space (e.g., spinning binaries). More importantly, 
several studies in the past have suggested that the contribution from subdominant modes are appreciable 
only for very massive binaries with large mass ratios~\cite{Pekowsky:2013hm,Brown:2013hm,Capano:2013hm,Varma:2014hm}. 
The effect of subdominant modes was thoroughly investigated in the context of GW150914, and 
the study concluded that the effect of subdominant modes is negligible in the detection and parameter 
estimation of this event~\cite{Abbott:2016wiq,LSC_2016nrsystematics}.

In a previous study~\cite{Varma:2014hm}, we investigated the effect of subdominant modes in the detection and parameter estimation of a population of nonspinning black hole binaries. Here, we extend our previous study to the case of black hole binaries with nonprecessing spins~\footnote{We note that, in a recent paper, Calderon-Bustillo \etal~\cite{CalderonBustillo:2016hm} extended our previous study of nonspinning binaries to the case of spinning binaries with equal component spins. Our new study covers a larger region in the parameter space, by employing numerical-relativity waveforms with larger mass ratios and spins. The template family that we use also can span a large spin range ($\chi_{1z,2z} \in [-1,1]$ as opposed to $\chi \in [-1,0.6]$ employed in \cite{CalderonBustillo:2016hm}); hence we see better fitting factors at the cost of a larger parameter bias.}. We construct target GW signals that include subdominant 
modes ($\ell \leq 4$, $m \neq 0$) by matching nonprecessing numerical-relativity simulations 
describing the late inspiral, merger and ringdown with post-Newtonian/effective-one-body 
waveforms describing the early inspiral. We then compute the reduction in the detectable 
volume (for a fixed SNR threshold) and systematic bias in the estimated parameters when 
nonprecessing quadrupole-mode only templates are employed in the detection and parameter 
estimation of these target waveforms. 

Figure~\ref{fig:summary_fig} summarizes the main results from this study. The left plot shows 
the region in the parameter space where neglecting subdominant modes will cause an 
unacceptable (more than 10\%) loss in the detectable volume (appropriately averaged over all
orientation and sky location angles) for a fixed SNR threshold. The right plot shows the region in the parameter space where 
neglecting subdominant modes will cause unacceptably large systematic bias in the parameter
estimation (i.e., systematic errors larger than the expected statistical errors for a sky and orientation-averaged SNR of 8). 
Comparing these results with our previous study employing nonspinning templates (i.e., by comparing the dashed green curve with the solid green curve in the left plot of Fig.~\ref{fig:summary_fig}), we see 
that including spin effects in the dominant-mode templates enhances their effectualness, 
thus reducing the region in 
the parameter space where subdominant mode templates are required for detection. 
However, this is achieved at the cost of introducing larger systematic errors in the 
estimated parameters, thus increasing the volume of the parameter space where
subdominant mode templates should be used in the parameter estimation. This effect 
(better effectualness at the cost of larger systematic errors) is more pronounced in the 
case of binaries with spins \emph{antialigned} with the orbital angular momentum. Thus, 
subdominant-mode templates are required for detection of binaries with antialigned spins only 
over a small region in the parameter space; but they are required for parameter estimation 
over a large region. This effect is reversed in the case of \emph{aligned} spins. 

The rest of this paper is organized as follows: Section~\ref{sec:methods} provides details 
of the methodology and figures of merit for this study. Section~\ref{sec:results} discusses 
our results including how we arrive at Fig.~\ref{fig:summary_fig}. 
Finally, Sec.~\ref{sec:conclusion} has some concluding remarks, limitations of this work and 
targets for future work. Appendix~\ref{sec:comparison_bayesian} presents a comparison of 
our estimates of the statistical and systematic errors with the same estimated from fully 
Bayesian parameter estimation for one sample case. Please note our notation for the rest of 
this article: $M$ refers to the total mass of the binary, $m_1$ and $m_2$ ($m_1 \geq m_2$) refer 
to the component masses, and $\chi_1$ and $\chi_2$ refer to the dimensionless spin parameters; 
$\chi_{1,2} = S_{1,2}/m_{1,2}^2$ where $S_{1,2}$ are the spin angular momenta of the components.  
All masses are detector frame (redshifted) masses. We only consider spins aligned/antialigned 
with the orbital angular momentum. The mass ratio is denoted by $q=m_1/m_2$ while 
$\eta=m_1 m_2/M^2$ denotes the symmetric mass ratio. We also define the effective spin 
parameters $\chieff = (m_1 \chi_1 + m_2 \chi_2)/M$ and 
$\mchieff = ({m_1 \chi_1 - m_2 \chi_2})/M$. We refer to waveforms that include contributions 
from sub-dominant modes ($\ell \leq 4$, $m \neq 0$) as ``full'' waveforms and waveforms that 
include only quadrupole modes ($\ell = 2, m = \pm 2$) as ``quadrupole'' waveforms. We refer to 
the SNR averaged over orientation and inclination angles as the orientation-averaged SNR; 
note that SNR along optimal orientation is $\sim2.5$ times the orientation-averaged 
SNR~\cite{2009lrr:SathyaSchutz}.

\section{Methodology}
\label{sec:methods}

\begin{table}
\centering
\begin{tabular}{c@{\quad}c@{\quad}c@{\quad}c@{\quad}c@{\quad}c@{\quad}c}
\toprule
Simulation ID & $q$ & $\chi_{1z}$ & $\chi_{2z}$ & $M\omega_\mathrm{orb}$ & Number of orbits\\
\midrule
SXS:BBH:0172 & $1$  &  $0.98$  & $0.98$  &  $0.015$ &  $25.4$\\   
SXS:BBH:0160 & $1$  &  $0.90$ & $0.90$ &  $0.015$   &  $24.8$\\ 
SXS:BBH:0155 & $1$  &  $0.80$ & $0.80$ &  $0.015$   &  $24.1$\\ 
SXS:BBH:0152 & $1$  &  $0.60$ & $0.60$ &  $0.016$   &  $22.6$\\ 
SXS:BBH:0090 & $1$  &  $0.00$  & $0.00$  &  $0.011$ &  $32.4$\\   
SXS:BBH:0151 & $1$  &  $-0.60$ & $-0.60$ &  $0.016$ &  $14.5$\\ 
SXS:BBH:0154 & $1$  &  $-0.80$ & $-0.80$ &  $0.016$ &  $13.2$\\ 
SXS:BBH:0159 & $1$  &  $-0.90$ & $-0.90$ &  $0.016$ &  $12.7$\\ 
SXS:BBH:0156 & $1$  &  $-0.95$ & $-0.95$ &  $0.016$ &  $12.4$\\   
SXS:BBH:0253 & $2$  &  $0.50$  & $0.50$  &  $0.014$ &  $28.8$\\   
SXS:BBH:0047 & $3$  &  $0.50$  & $0.50$  &  $0.017$ &  $22.7$\\   
SXS:BBH:0174 & $3$  &  $0.50$  & $0.00$  &  $0.013$ &  $35.5$\\   
SXS:BBH:0110 & $5$  &  $0.50$  & $0.00$  &  $0.019$ &  $24.2$\\   
SXS:BBH:0202 & $7$  &  $0.60$  & $0.00$  &  $0.013$ &  $62.1$\\   
SXS:BBH:0203 & $7$  &  $0.40$  & $0.00$  &  $0.013$ &  $58.5$\\   
SXS:BBH:0065 & $8$  &  $0.50$  & $0.00$  &  $0.019$ &  $34.0$\\   
SXS:BBH:0184 & $2$  &  $0.00$  & $0.00$  &  $0.018$ &  $15.6$\\   
SXS:BBH:0183 & $3$  &  $0.00$  & $0.00$  &  $0.020$ &  $15.6$\\   
SXS:BBH:0167 & $4$  &  $0.00$  & $0.00$  &  $0.021$ &  $15.6$\\   
SXS:BBH:0056 & $5$  &  $0.00$  & $0.00$  &  $0.016$ &  $28.8$\\   
SXS:BBH:0181 & $6$  &  $0.00$  & $0.00$  &  $0.018$ &  $26.5$\\   
SXS:BBH:0298 & $7$  &  $0.00$  & $0.00$  &  $0.021$ &  $19.7$\\   
SXS:BBH:0063 & $8$  &  $0.00$  & $0.00$  &  $0.019$ &  $25.8$\\   
SXS:BBH:0189 & $9.2$  &  $0.00$  & $0.00$  &  $0.021$ &  $25.2$\\ 
SXS:BBH:0185 & $10$  &  $0.00$ & $0.00$  &  $0.021$ &  $24.9$\\   
SXS:BBH:0238 & $2$  &  $-0.50$ & $-0.50$ &  $0.011$ &  $32.0$\\   
SXS:BBH:0036 & $3$  &  $-0.50$ & $0.00$  &  $0.012$ &  $31.7$\\   
SXS:BBH:0046 & $3$  &  $-0.50$ & $-0.50$ &  $0.018$ &  $14.4$\\   
SXS:BBH:0109 & $5$  &  $-0.50$ & $0.00$  &  $0.020$ &  $14.7$\\   
SXS:BBH:0205 & $7$  &  $-0.40$ & $0.00$  &  $0.013$ &  $44.9$\\   
SXS:BBH:0207 & $7$  &  $-0.60$ & $0.00$  &  $0.014$ &  $36.1$\\   
SXS:BBH:0064 & $8$  &  $-0.50$ & $0.00$  &  $0.020$ &  $19.2$\\   
\bottomrule
\end{tabular} 
\caption{Summary of the parameters of the NR waveforms used in this paper: $q \equiv m_1/m_2$ is the mass ratio of the binary, $\chi_{1z}$ and $\chi_{2z}$ are the dimensionless spins of the larger and smaller black holes respectively, and $M \omega_\mathrm{orb}$ is the orbital frequency after the junk radiation. All of these waveforms have residual eccentricity, $e<4\times10^{-3}$ (typically significantly smaller).}
\label{tab:nr_waveforms}
\end{table}

\begin{figure}[hbt]
\begin{center}
\includegraphics[scale=0.4]{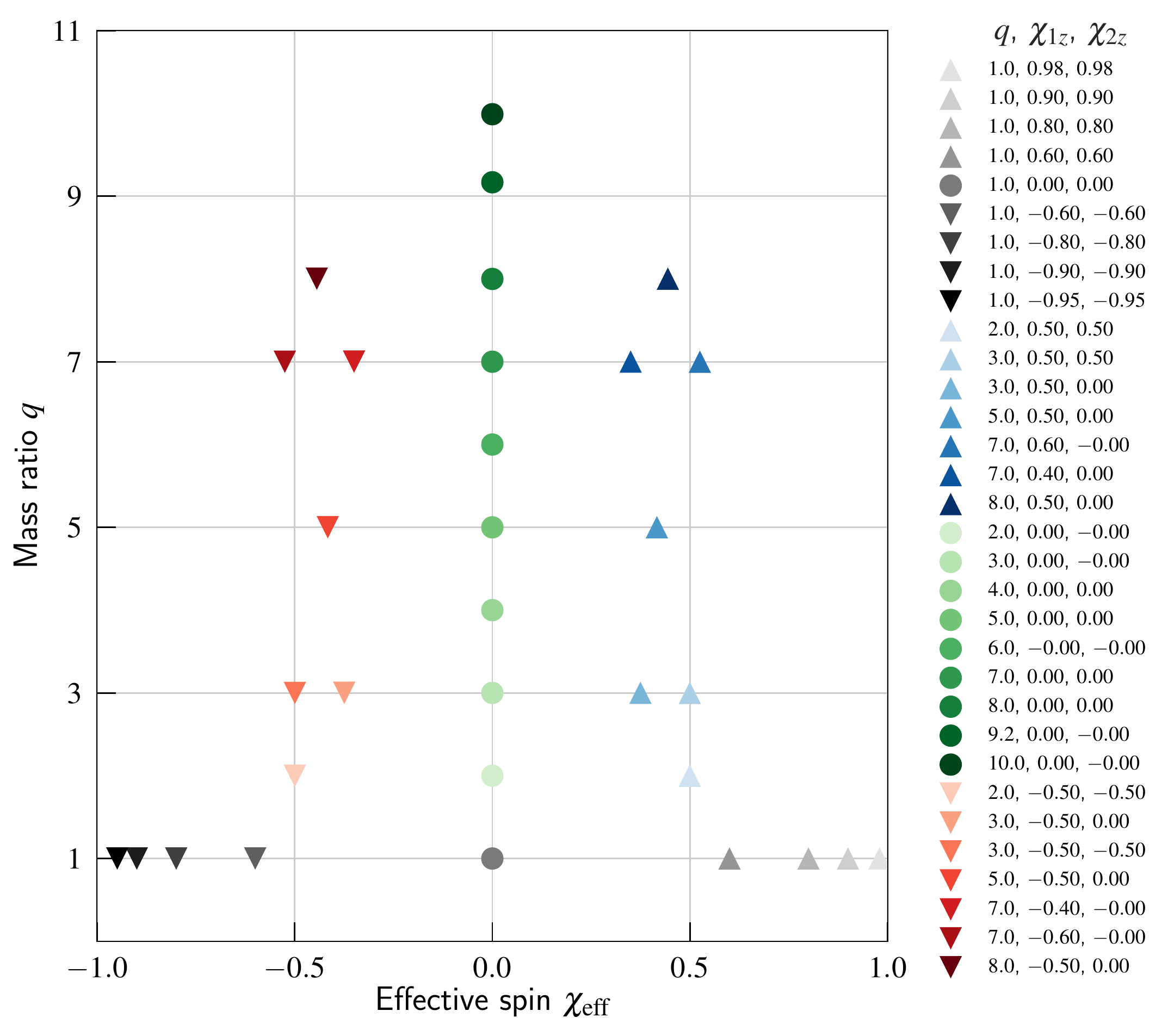}
\caption{This plot shows the mass ratio (vertical axis) and effective spin (horizontal axis) of the NR waveforms used in this study. The color scheme of the markers is same as that in Figs.~\ref{fig:summary_fig}, \ref{fig:effparams} and \ref{fig:minsnr}, enabling direct comparison.}
\label{fig:data_paramspace}
\end{center}
\end{figure}

In a past study~\cite{Varma:2014hm}, we investigated the effects of nonquadrupole modes 
in the detection and parameter estimation of nonspinning binaries. Here we extend the 
earlier work to the case of nonprecessing binaries, covering a wide range of total 
masses ($40\Mo\leq M\leq300\Mo$), mass ratios ($q\leq10$) and 
spins ($- 0.5 \lesssim \chieff \lesssim 0.5$ for 
$q\geq2$ and  $-0.95 \leq \chieff \leq 0.98$ for $q=1$).  
For our target signals, we use hybrid waveforms constructed by matching NR waveforms 
that describe the late inspiral, merger and ringdown of binary black holes with 
post-Newtonian (PN) / effective-one-body (EOB) waveforms modeling the early inspiral. 
These hybrids contain several nonquadrupolar modes ($\hlm(t)$ with 
$\ell \leq 4, |m| \leq \ell, m \neq 0$) of GW signals from binary black holes. 
The PN waveforms were generated using the 3PN amplitude given 
by~\cite{Blanchet:2008je,Arun:2009,Buonanno:2012rv}, but using the phase evolution 
given by the SEOBNRv2 waveform family\footnote{This was done in order to make the 
phase evolution of the hybrids very similar to that of the templates, so that a mismatch 
between the hybrid and the template due to the different phase evolution will not be 
mistaken as due to the effect of subdominant modes.}~\cite{PhysRevD.89.061502}. 
We match them with NR waveforms produced by the 
SpEC~\cite{SXS-Catalog, SpEC, Mroue:2013xna, Mroue:2012kv, Lovelace:2010ne,  
Blackman:2015pia,   Buchman:2012dw, Ossokine:2013zga, Hemberger:2012jz, Szilagyi:2009qz, 
Boyle:2009vi, Scheel:2008rj, Boyle:2007ft, Scheel:2006gg, Lindblom:2005qh, 
Pfeiffer:2002wt,Hemberger:2013hsa} code by the SXS Collaboration that are available 
at the public SXS catalog of NR waveforms~\cite{SXS-Catalog}. The parameters of the NR 
waveforms used in this study are shown in Table~\ref{tab:nr_waveforms} and 
Fig.~\ref{fig:data_paramspace}. Note that the ($\ell,m$) = (4,1) mode in several of 
the NR waveforms has significant numerical noise. However, as the amplitude of this mode 
is several orders of magnitude smaller than that of the dominant mode, we do not expect this 
to impact our results.

\begin{figure*}[htb]
\begin{center}
\includegraphics[width=7in]{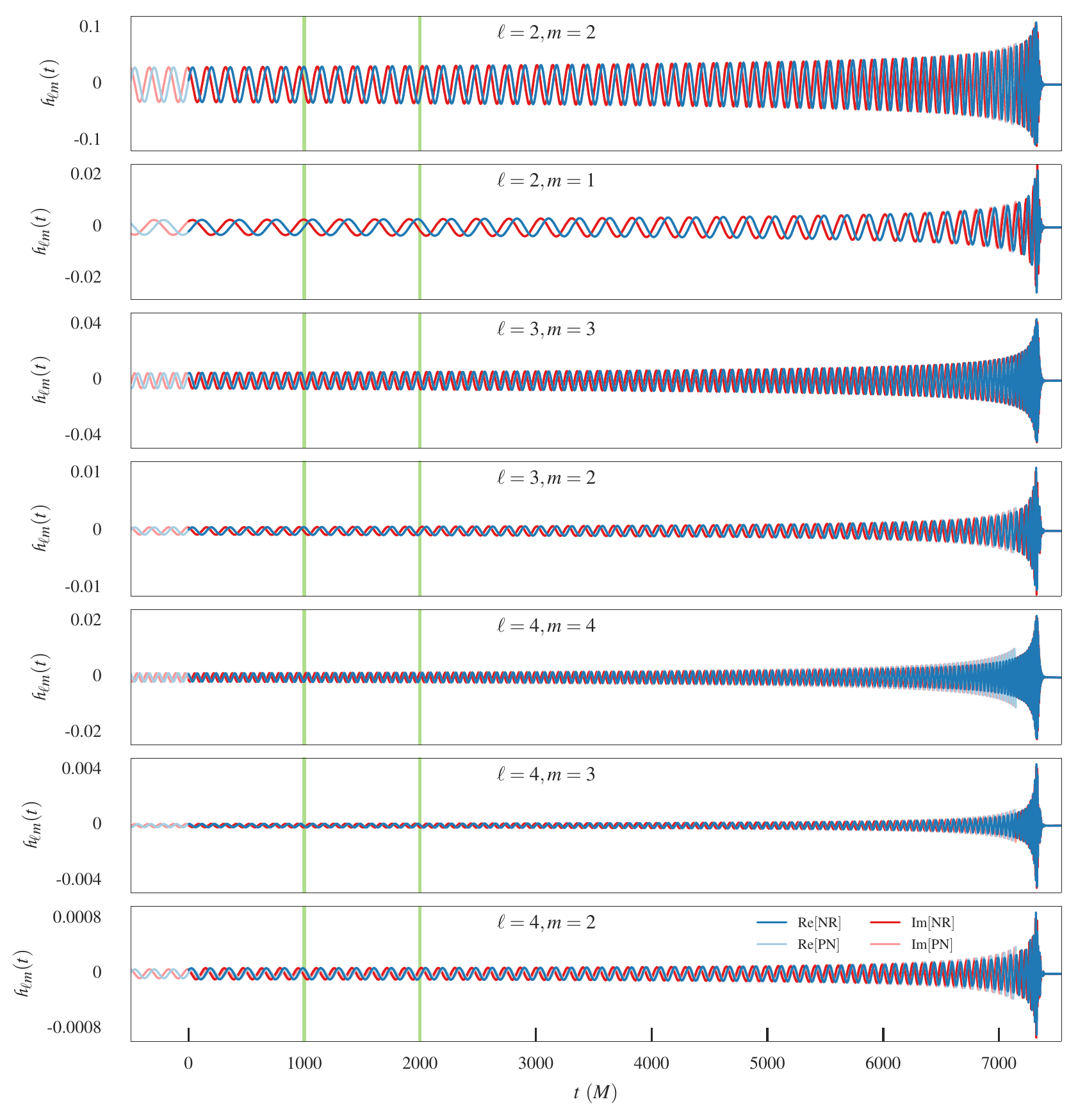}
\caption{Example of hybrid waveform modes constructed by matching NR and PN modes. These hybrid waveforms are constructed by matching $q=8$, $\chi_{1z}=0.5$, $\chi_{2z}=0$ NR waveforms computed using the SpEC code with PN/EOB waveforms describing the early inspiral. The horizontal axes show the time (with origin at the start of the NR waveforms) and the vertical axes show the GW modes $\hlm(t)$. The matching region $(1000 M, 2000 M)$ is marked by vertical green lines.}
\label{fig:hybrid_modes}
\end{center}
\end{figure*}

As described in detail in our past study~\cite{Varma:2014hm}, to construct hybrids, we match the PN modes $\hlm^\mathrm{PN}(t)$ with NR modes $\hlm^\mathrm{NR}(t)$ by a least square fit over two rotations ($\varphi_0, \psi$) on the NR mode and the time difference between NR and PN modes:
\begin{equation}
\Delta = \mathrm{min}_{t_0,\varphi_0, \psi} \int_{t_1}^{t_2} dt \sum_{\ell, m} \left|\hlm^\mathrm{NR}(t-t_0) \mathrm{e}^{i (m \varphi_0 + \psi)}  - \hlm^\mathrm{PN}(t) \, \right|.
\end{equation}
The hybrid modes are constructed by combining the NR modes with the ``best-matched'' PN modes:
\begin{equation}
\hlm^\mathrm{hyb}(t) \equiv \, \tau(t) \, \hlm^\mathrm{NR}(t-t_0') \ e^{i(m\varphi_0'+\psi')} + (1-\tau(t)) \, \hlm^\mathrm{PN}(t),
\end{equation}
where $t_0', \varphi_0'$ and $\psi'$ are the values of $t_0, \varphi_0$ and $\psi$ that 
minimizes the difference $\Delta$ between PN and NR modes and $\tau(t)$ is a suitable 
weighting function that smoothly goes from 0 to 1 during the interval 
$t_1 \leq t \leq t_2$. We refer the reader to Ref.~\cite{Varma:2014hm} for details about the
construction of hybrid waveforms. An example of hybrid waveform modes is shown in 
Fig.~\ref{fig:hybrid_modes}. It can be seen that higher modes are excited only during 
the very late inspiral, merger and ringdown. The effect of higher modes will be appreciable 
only in the mass range where the SNR contributed by the merger-ringdown is a significant 
fraction of the total SNR. This is the reason we restrict our study to the mass 
range $40 M_\odot \leq M \leq 300 M_\odot$; we do not see any evidence of a significant 
impact of higher modes for binaries with lower masses. Since the NR waveforms we use 
include tens of cycles in the inspiral, we do not expect hybridization errors to impact 
our results, particularly for high masses. For a detailed study on hybridization errors, 
we refer the reader to 
Refs.~\cite{MacDonald:2012mp, MacDonald:2011ne, Hannam:2007ik, Ajith:2007xh, Hannam:2010ky}.

\begin{figure*}[h!bt]
\captionsetup[subfigure]{labelformat=empty, singlelinecheck=true}
\begin{center}
    \subfloat[]
    {\includegraphics[scale=\skymapscale]{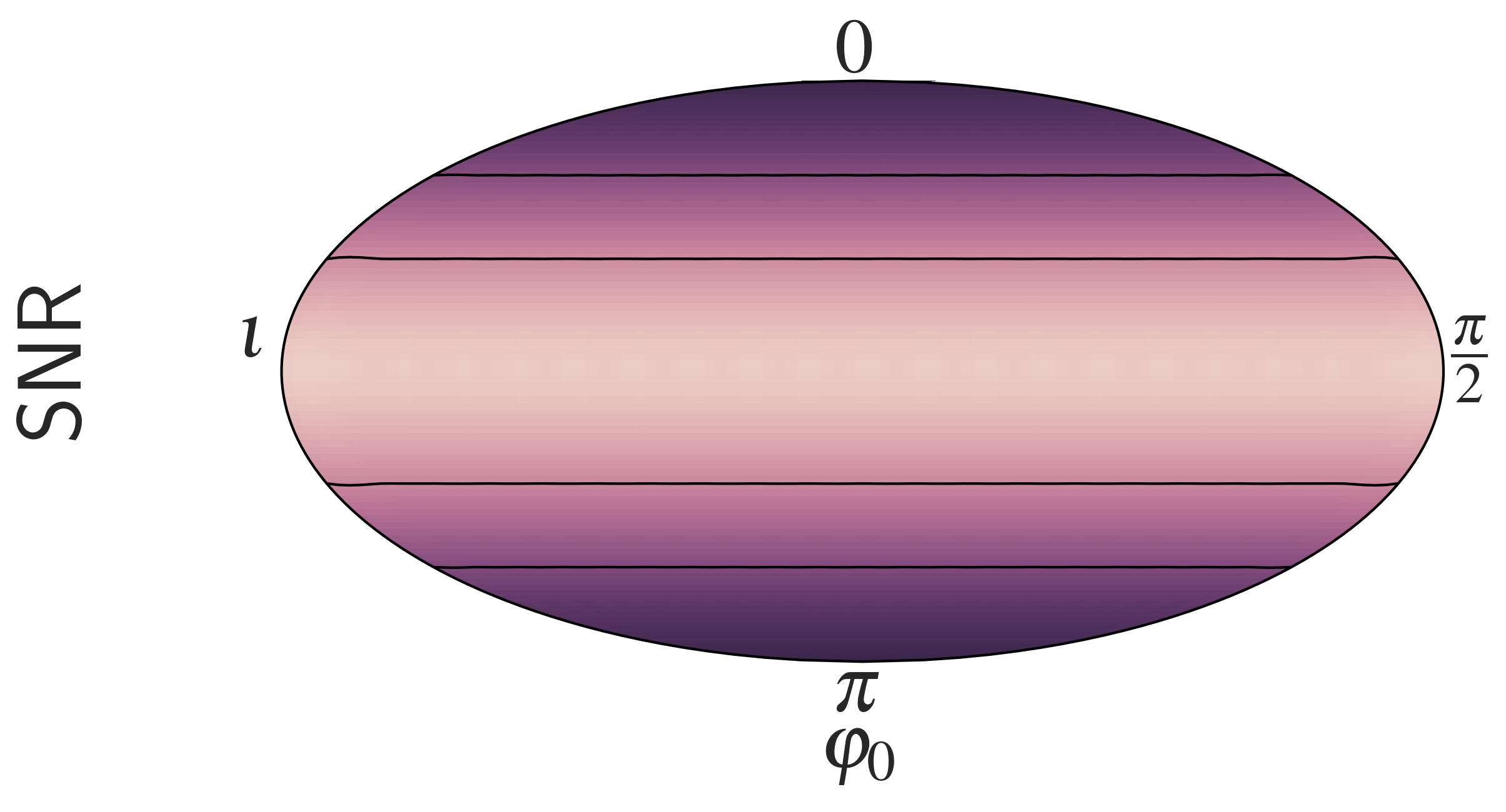}}
    \subfloat[]
    {\includegraphics[scale=\skymapscale]{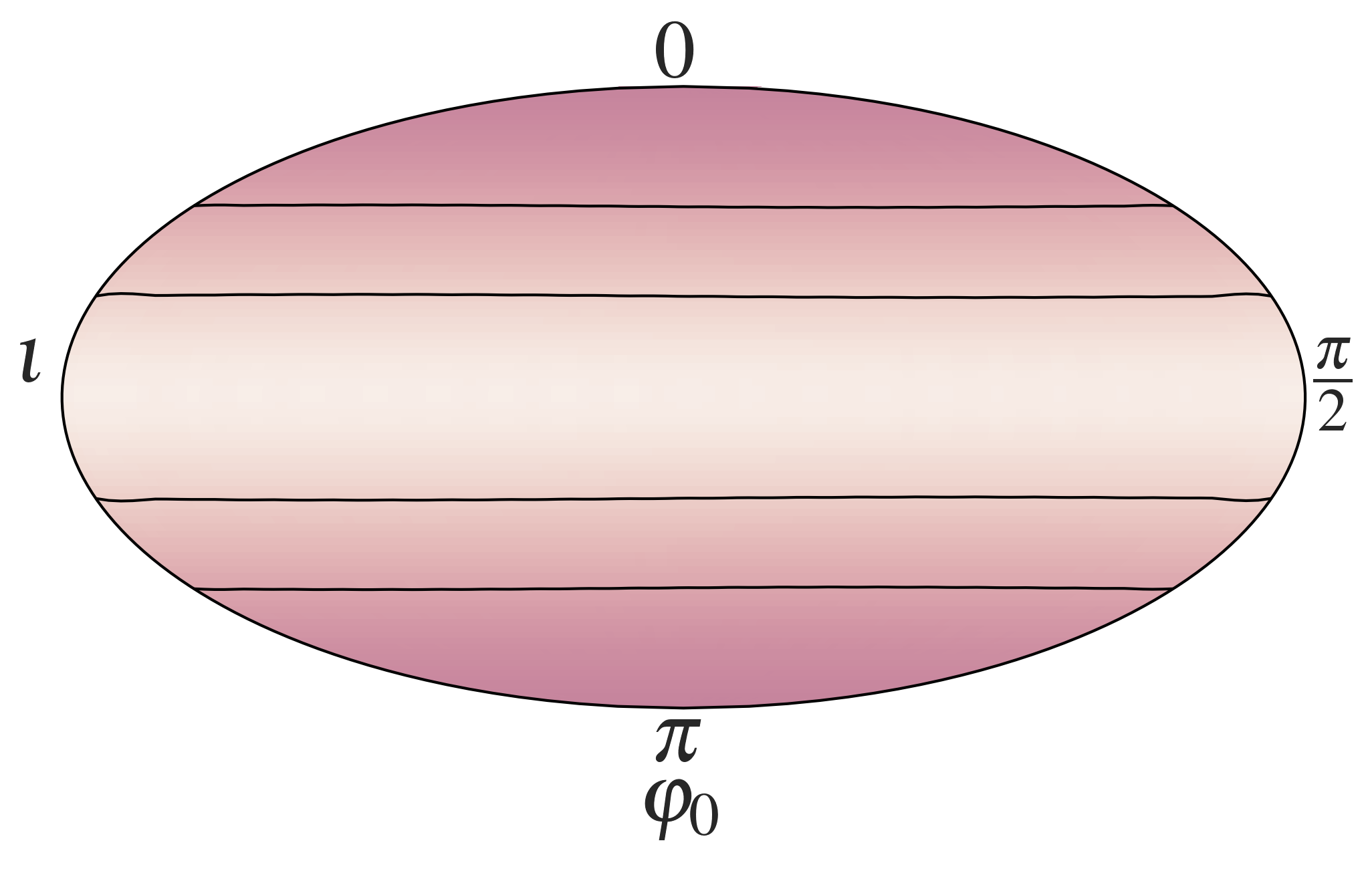}}
    \subfloat[]
    {\includegraphics[scale=\skymapscale]{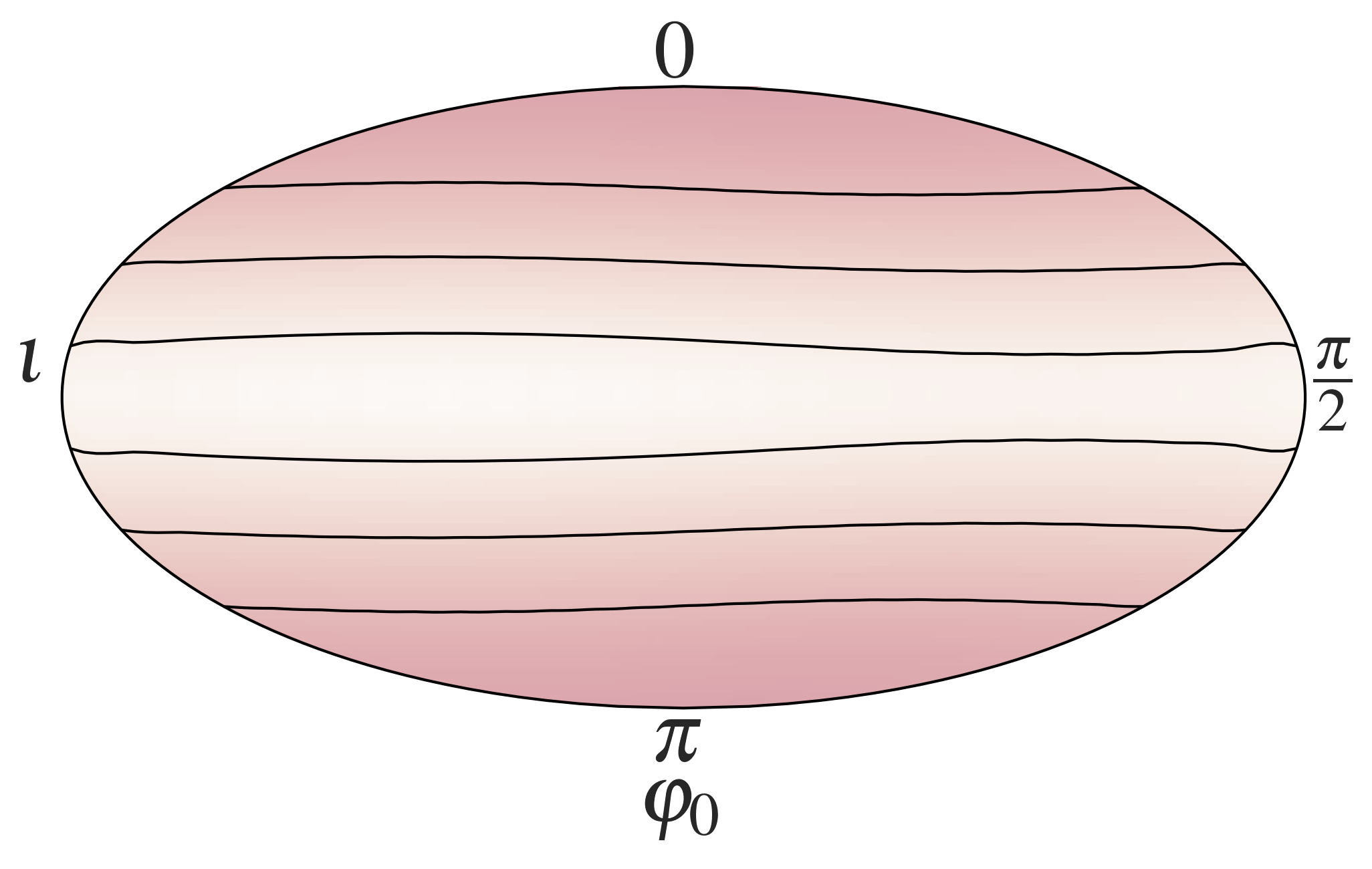}}
    \subfloat[]
    {\includegraphics[scale=\skymapscale]{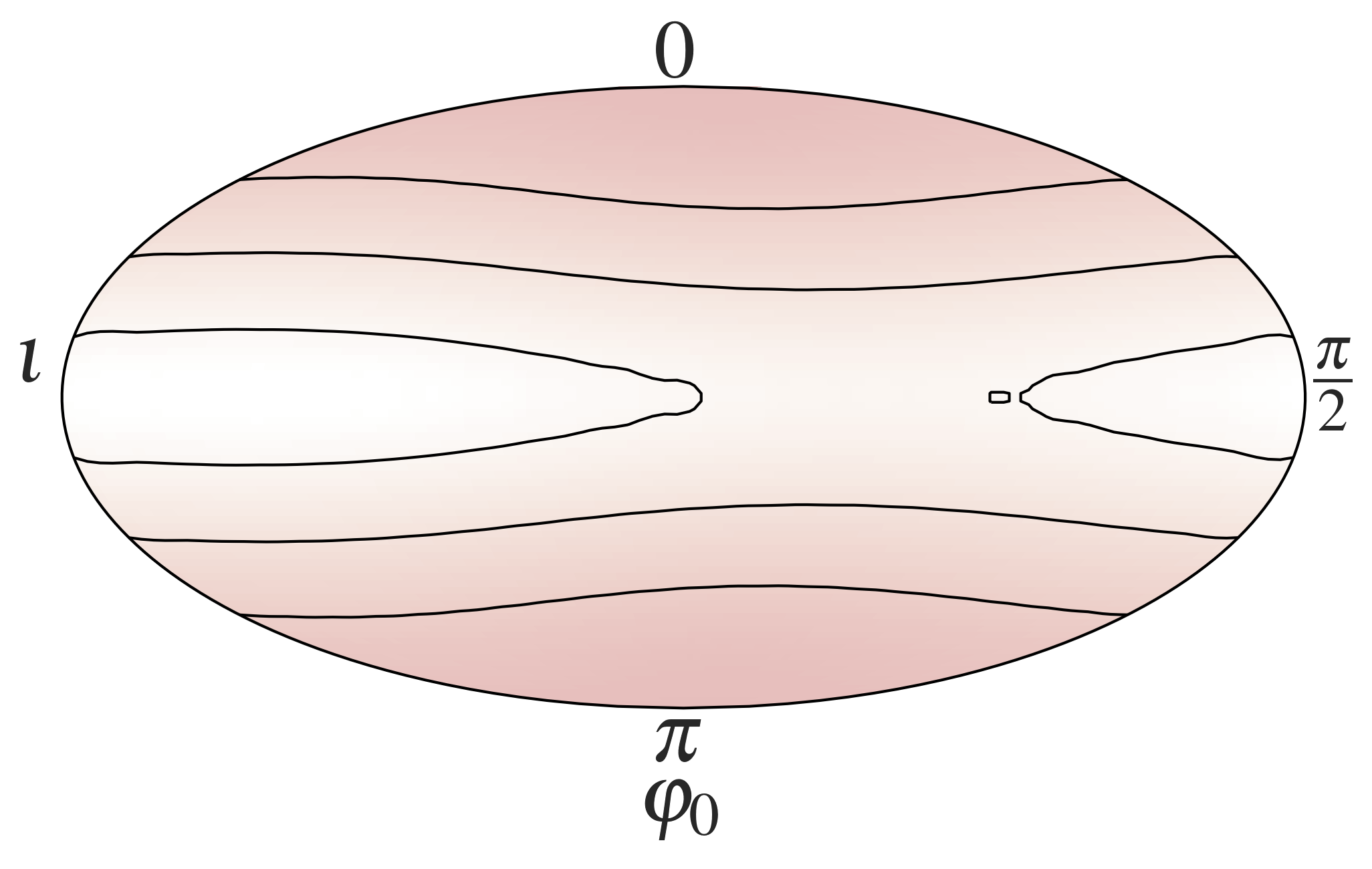}}
    {\includegraphics[scale=0.6]{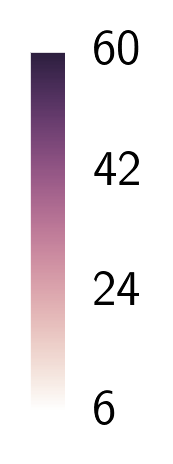}}

    \subfloat[$~q=1$, $\chi_{1z}=0.0$\label{subfig:skymap_ff_q1_s0}]
    {\includegraphics[scale=\skymapscale]{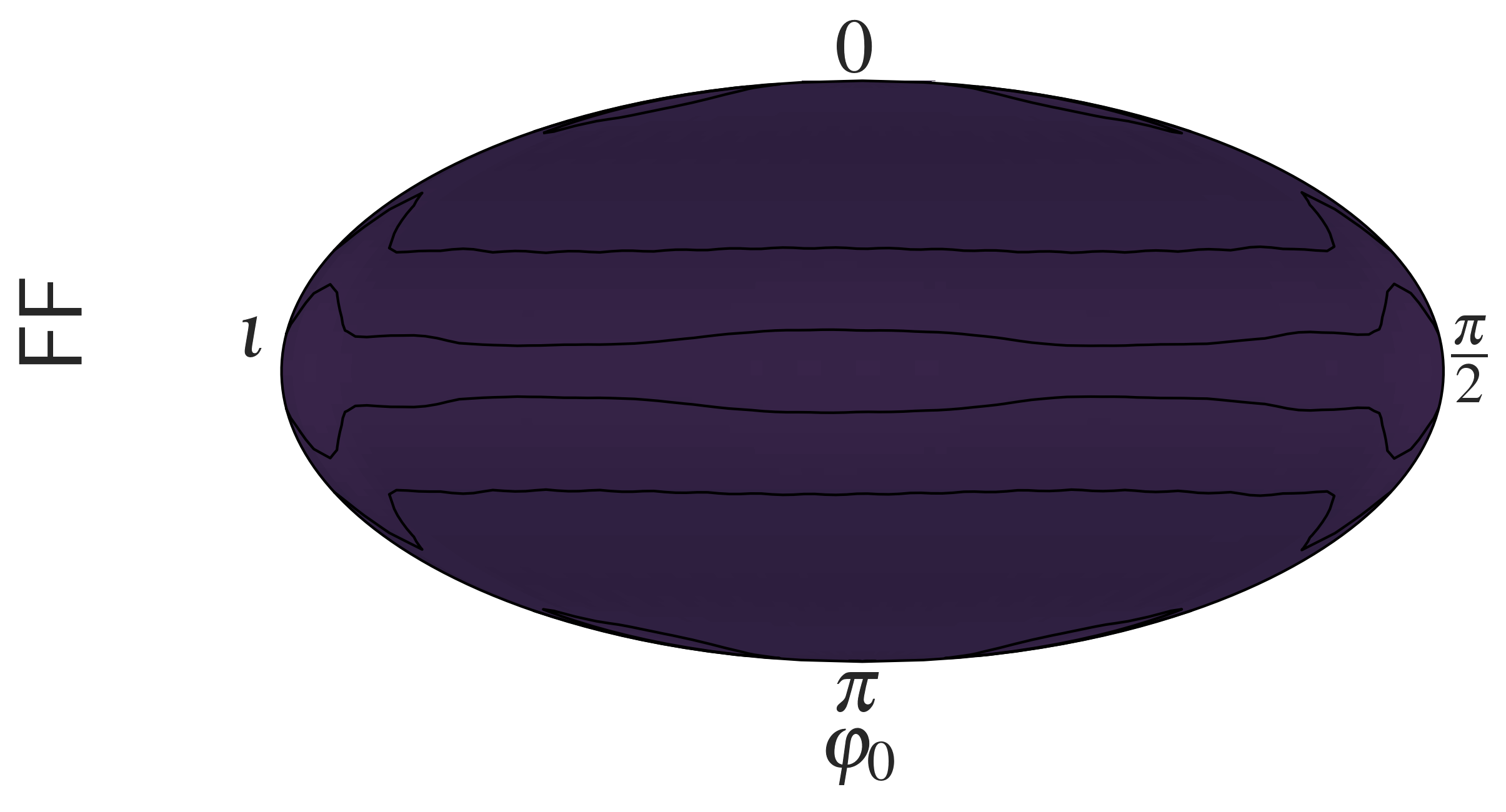}}
    \subfloat[$~q=8$, $\chi_{1z}=0.5$\label{subfig:skymap_ff_q8_s0.5}]
    {\includegraphics[scale=\skymapscale]{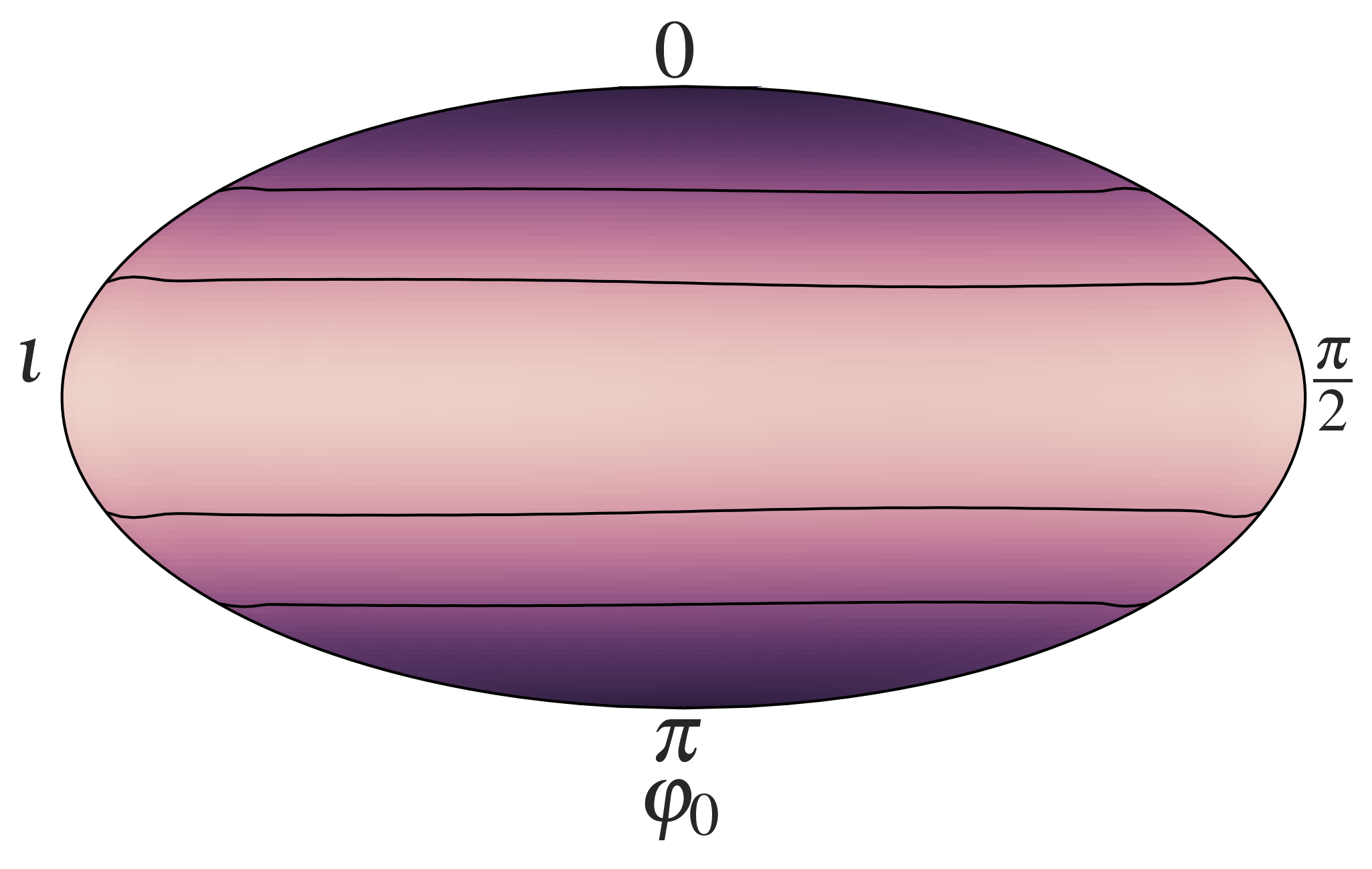}}
    \subfloat[$~q=8$, $\chi_{1z}=0.0$\label{subfig:skymap_ff_q8_s0}]
    {\includegraphics[scale=\skymapscale]{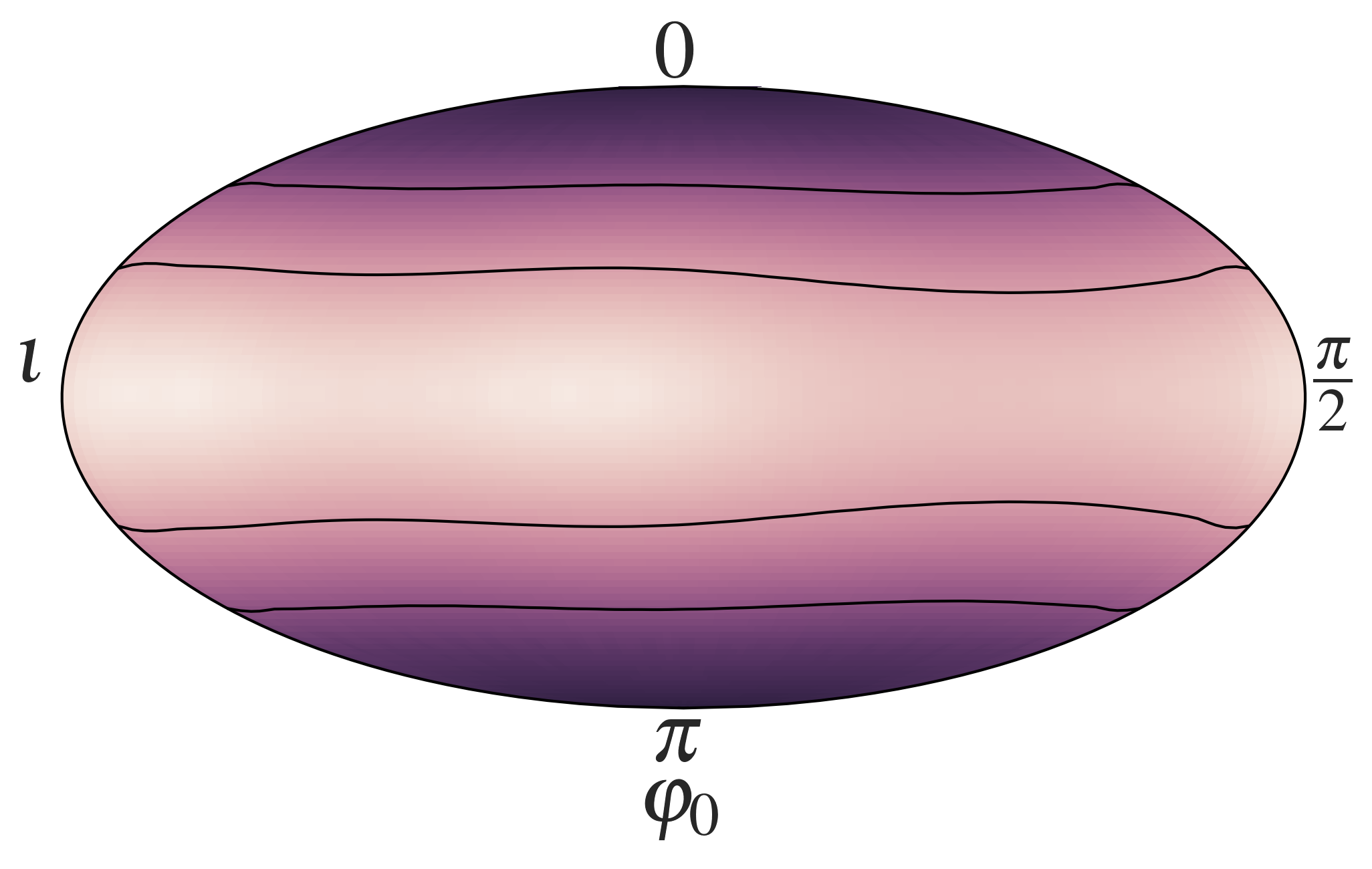}}
    \subfloat[$~q=8$, $\chi_{1z}=-0.5$\label{subfig:skymap_ff_q8_s-0.5}]
    {\includegraphics[scale=\skymapscale]{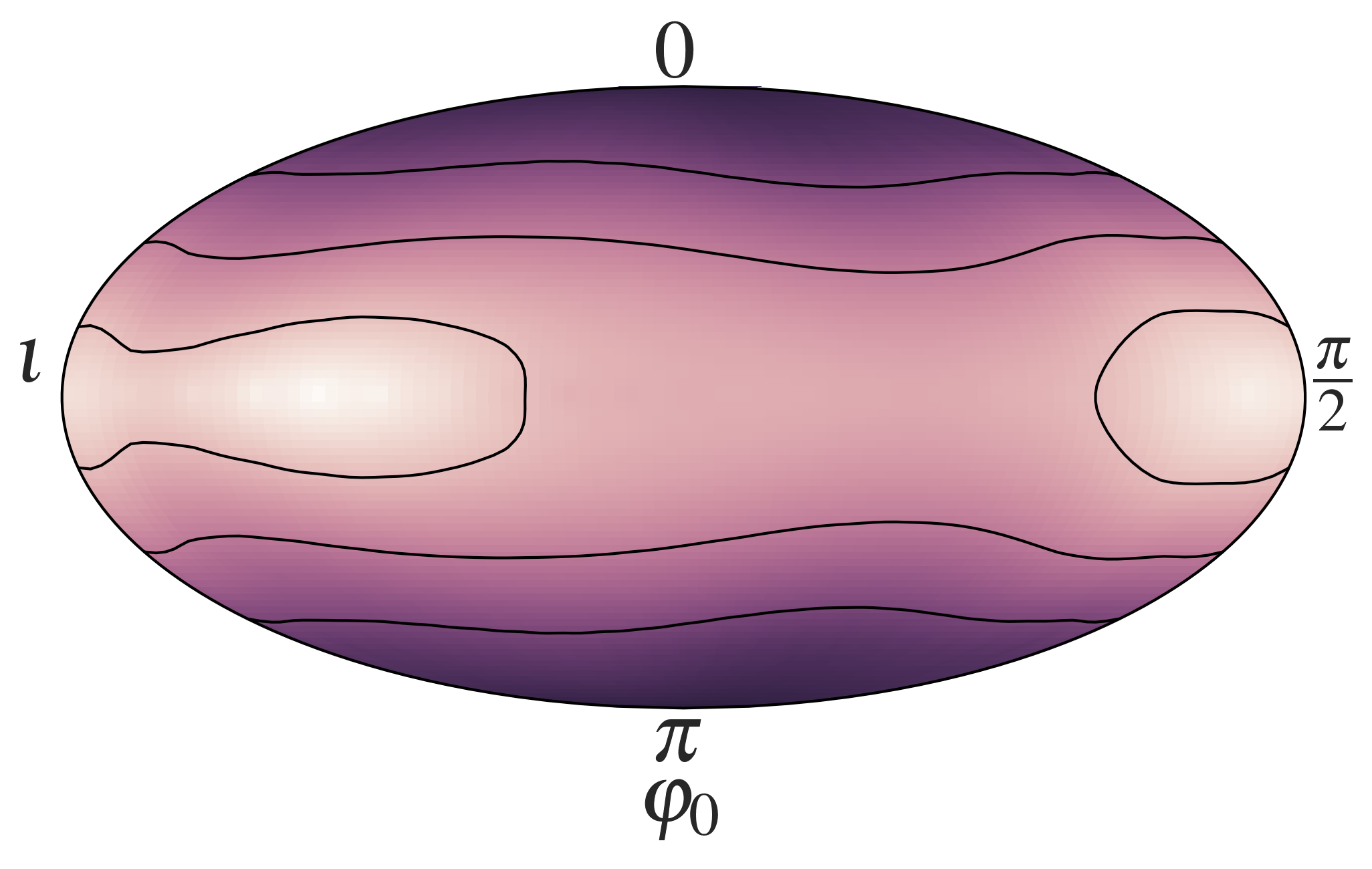}}
    {\includegraphics[scale=0.6]{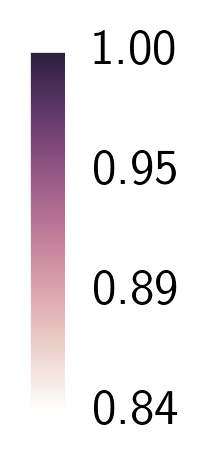}}
\caption{Optimal SNR (top panel) and fitting factor of quadrupole templates (bottom panel), 
averaged over polarization angle $\psi$ for binaries with total mass $M=100~\Mo$, 
located at 1 Gpc. The y-axis shows the inclination angle $\iota$ in radians and the 
x-axis shows the initial phase of the binary $\varphi_0$ in radians. 
The equator ($\iota=\pi/2$) corresponds to ``edge-on'' orientation while the 
poles ($\iota=0,\pi$) correspond to ``face-on'' orientation. Different columns correspond 
to different mass ratios and spins of the larger black hole (the spin on the smaller black 
hole is 0 in all three cases). It may be noted that the fitting factor as well as the intrinsic 
luminosity are smallest (largest) at $\iota = \pi/2 ~(\iota = 0,\pi)$ where contribution 
from the nonquadrupolar modes is the largest (smallest), illustrating the selection bias 
toward configurations where nonquadrupole modes are less important.}
\label{fig:skymap_ff_snr}
\end{center}
\end{figure*}

\begin{figure*}[tbh]
\begin{centering}
\includegraphics[scale=0.45]{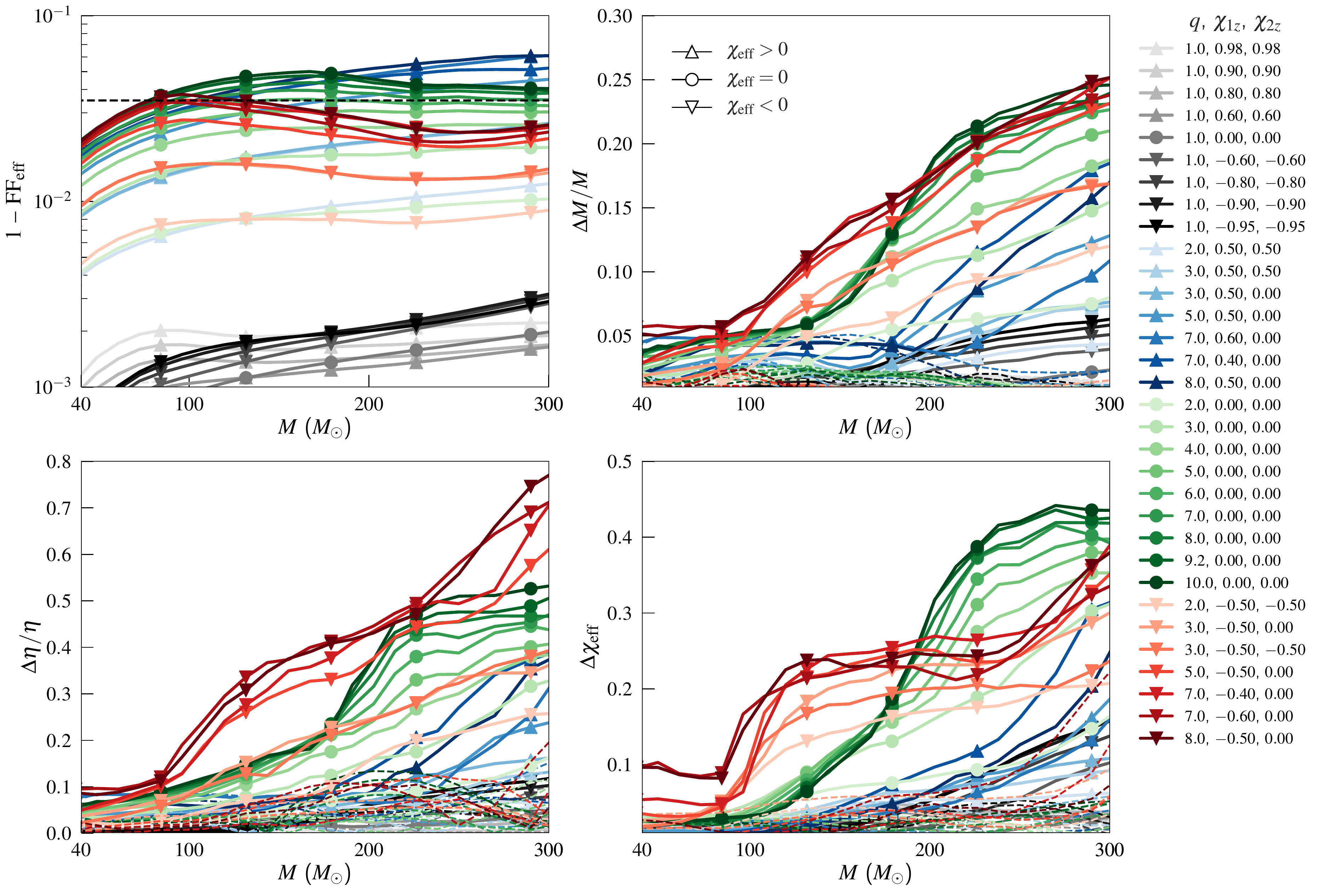}
\caption{``Ineffectualness'' (1 - $\FFe$) and effective parameter biases when using quadrupole-mode templates against hybrid waveforms including all modes. Dashed lines correspond to the same but against quadrupole-only hybrid waveforms, so that the difference between the dashed and sold lines gives an indication of the effect of nonquadrupole modes. Fractional biases are shown for total mass $M$ and symmetric mass ratio $\eta$, while absolute biases are shown for effective spins $\chieff$. $\FFe$ and effective parameter biases are obtained by averaging over all relevant orientations of the binary using Eqs.~(\ref{eq:efftive_vol}) and (\ref{eq:avg_syst_err}). The horizontal axis reports the total mass of the binary while the mass ratio and spins are shown in the legend. The markers indicate the spin types: triangles pointing up/down denoting binaries with aligned/antialigned spins and circles denoting nonspinning binaries. 
The horizontal dashed black line corresponds to $1-\FFe^3 = 0.1$. Note that most of the dashed lines in the top-left subplot lie below $10^{-3}$. We see that as the total mass increases, the ineffectualness and effective biases in $M$, $\eta$ and $\chieff$ increase and are dominated by the effects of subdominant modes; see Sec.~\ref{sec:results} for further discussion.}
\label{fig:effparams}
\end{centering}
\end{figure*}

The template family used is \TemplateName~\cite{Khan_2016IMRPhenomD, Husa_2016IMRPhenomD}, which is a quadrupole-only ($\ell=2, m=\pm2$) inspiral, merger and ringdown waveform family described by two mass parameters and two nonprecessing spin parameters. These waveforms are calibrated to NR waveforms with $q \leq 18$, $|\chi_{1z,2z}| \lesssim 0.85$ ($0.98$ for $q=1$)  and we find that they 
have a very good agreement with the quadrupole modes of the hybrid waveforms discussed above (cf. the dashed lines in Fig.~\ref{fig:effparams}). The waveforms are generated in the Fourier-domain using the LALSimulation~\cite{lalsimulation} software package. 

We compute fitting factors~\cite{Apostolatos:1995pj} by maximizing the overlap (noise 
weighted inner product) of the template family against the target hybrid signals and infer 
the systematic errors by comparing the best-match parameters with the true parameters. 
The overlaps are maximized over the extrinsic parameters (time of arrival $t_0$ and the 
reference phase $\varphi_0$) using the standard techniques in GW data analysis 
(see, e.g., Ref.~\cite{Allen:2005fk}), while the overlaps are maximized over the 
intrinsic parameters ($M$,  $\eta$, $\chi_{1z}$ and $\chi_{2z}$) of the templates using 
a Nelder-Mead downhill simplex algorithm~\cite{scipy}, with additional enhancements described 
in Ref.~\cite{Varma:2014hm}. For the model of the noise power spectrum, we use the 
``zero-detuned, high-power" design noise power spectral density (PSD)~\cite{adligo-psd} of Advanced LIGO with 
a low frequency cut-off of 20 Hz. 

The contribution of subdominant modes in the observed signal depends on the relative 
orientation of the binary and the detector. The SNR (and hence the volume in the local 
Universe where the binary can be confidently detected) is also a strong function of this 
relative orientation. For, e.g., binaries that are face-on produce the largest SNR in 
the detector; however, the contribution from subdominant modes is minimal for this orientation. 
This effect is reversed for the case of edge-on orientations. Thus, if we want to calculate 
the effect of subdominant modes on detection and parameter estimation of a population of 
binary black holes, the effect has to be averaged over all orientations after appropriately 
weighting each orientation. 

We evaluate the \emph{effective volume}~\cite{Varma:2014hm} of a search, defined as the fraction of the volume that is accessible by an optimal search (corresponding to a fixed SNR threshold), by averaging over all the relative orientations in the following way: 
\begin{equation}
\label{eq:efftive_vol}
V_\mathrm{eff}\,(m_1,m_2,\chi_{1z},\chi_{2z}) = \frac{\overline{\rhoopt^3 \, \FF^3}}{\overline{\rhoopt^3}},
\end{equation}
where $\rhoopt$ is the optimal SNR of the full signal, $\FF$ is the fitting factor of the dominant mode template, and the bars indicate averages over all (isotropically distributed) 
orientations~\footnote{This corresponds to uniform distributions in the phase angle 
$\varphi_0 \in [0, 2\pi)$, polarization angle $\psi \in [0, 2\pi)$, and the cosine of the 
inclination angle $\cos \iota \in [-1,1]$. Note that we assume that the binaries are optimally 
located (i.e., the angles $\theta,\phi$ describing the location of the binary in the detector 
frame on the sky are set to zero). The error introduced by this restriction is very small 
($\sim 0.1\%$) due to the weak dependence of the matches on $(\theta, \phi)$ and the strong 
selection bias towards binaries with $\theta \simeq 0, \pi$, where the antenna pattern 
function peaks~\cite{Varma:2014hm}. \label{Ftnote:angles}}. 
The dominant-mode template family is deemed effectual for detection when the effective volume is greater than 90\%; or when the \emph{effective fitting factor} $\FFe := V_\mathrm{eff}^{1/3}$ is greater than 0.965.

Similarly, we define the \emph{effective bias}~\cite{Varma:2014hm} in estimating an 
intrinsic parameter $\lambda$ as 
\begin{equation}
\label{eq:avg_syst_err}
\Delta \lambda_\mathrm{eff}(m_1,m_2,\chi_{1z},\chi_{2z}) = \frac{ \overline{ |\Delta \lambda| ~ \rhoopt^3 ~ \FF^3 } } {\overline{ \rhoopt^3 ~ \FF^3 } },
\end{equation}
where $\Delta \lambda$ is the systematic bias in estimating the parameter $\lambda$ for 
one orientation, $\FF$ is the corresponding fitting factor, and $\rhoopt$ the corresponding 
optimal SNR. Here, also, the bars indicate averages over all orientations. The effective bias 
provides an estimate of the bias averaged over a population of detectable binaries with 
isotropic orientations. We compare them against the sky and orientation averaged statistical 
errors. Statistical errors are computed using the Fisher matrix formalism employing 
quadrupole-only templates. The quadrupole-mode template family is deemed faithful for parameter 
estimation when the effective biases in all of the three intrinsic parameters 
$M, \eta, \chieff$ are smaller than the $1\sigma$ statistical errors in measuring the same 
parameter for an orientation-averaged SNR of 8.

\begin{figure*}[h!bt]
\captionsetup[subfigure]{labelformat=empty, singlelinecheck=true}
\begin{center}

    \subfloat[]
    {\includegraphics[scale=\skymapscale]{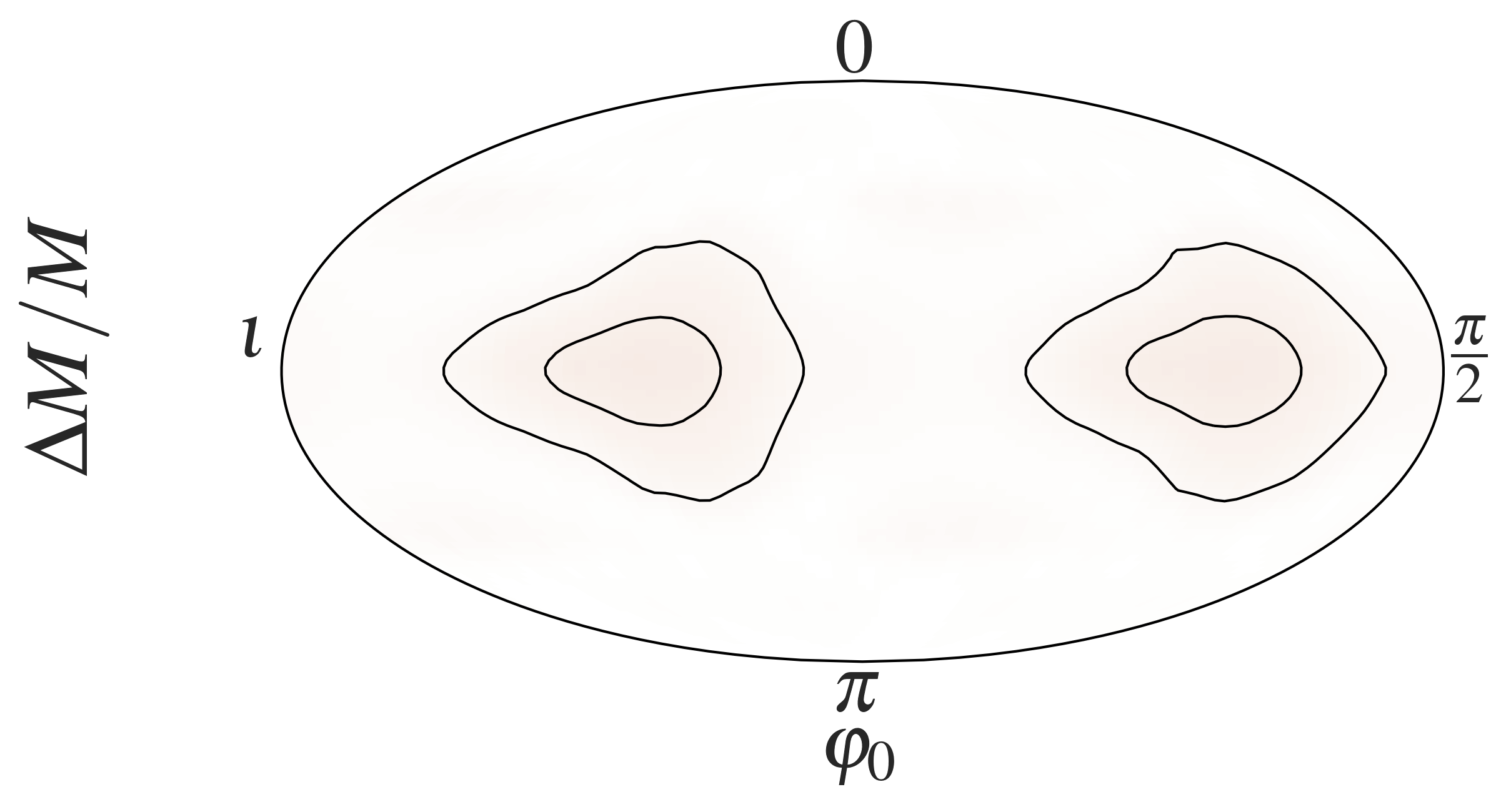}}
    \subfloat[]
    {\includegraphics[scale=\skymapscale]{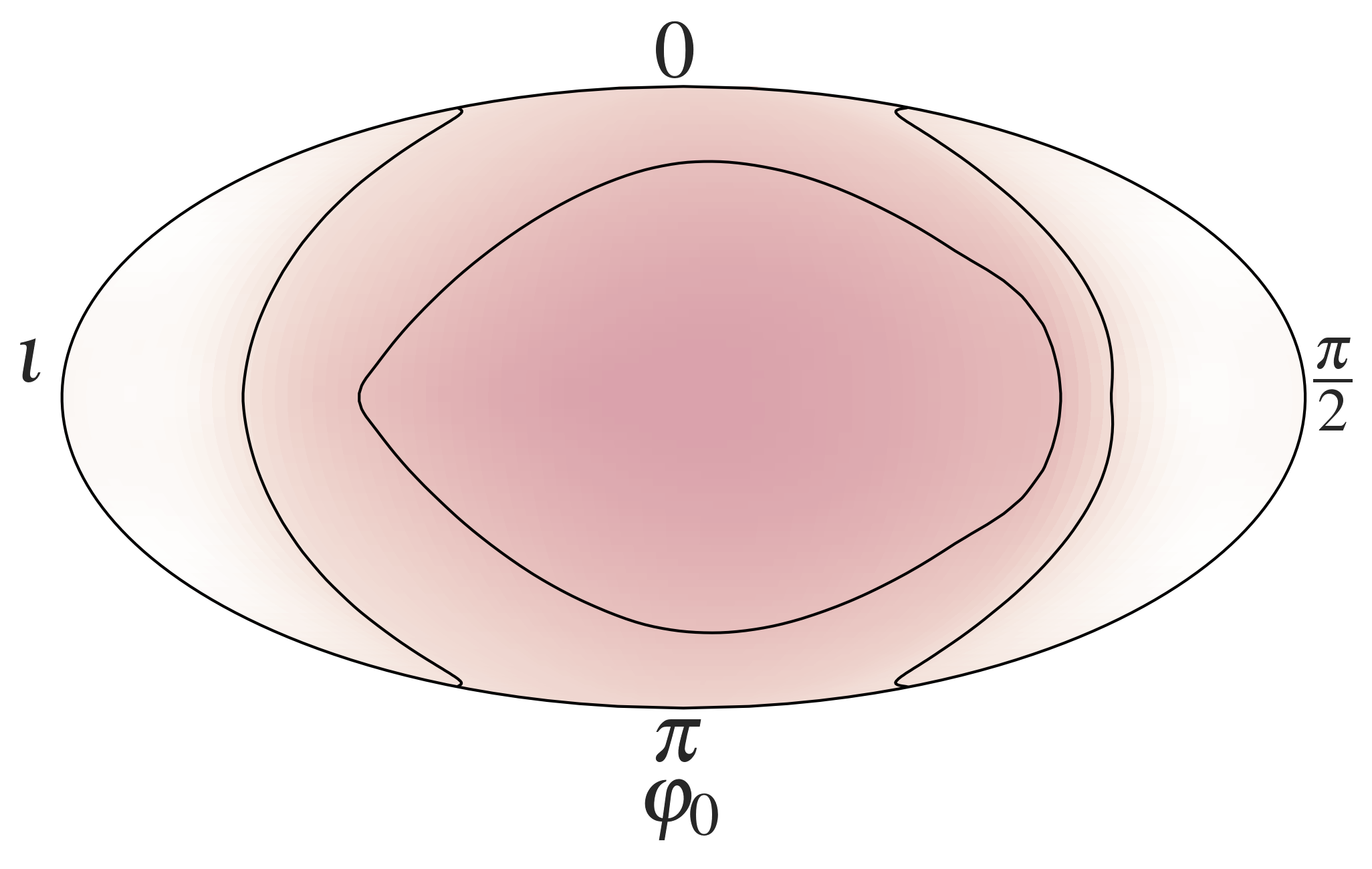}}
    \subfloat[]
    {\includegraphics[scale=\skymapscale]{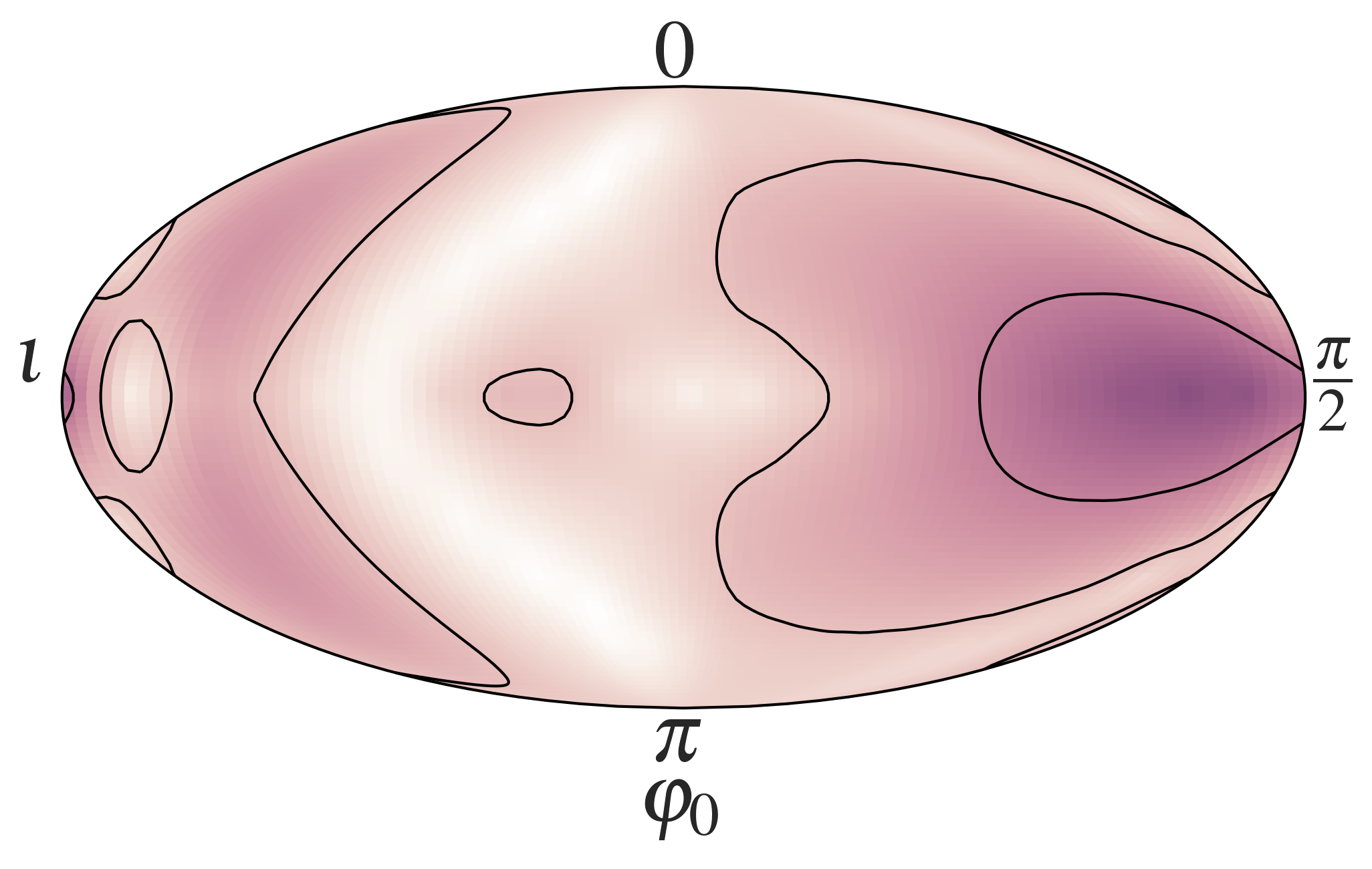}}
    \subfloat[]
    {\includegraphics[scale=\skymapscale]{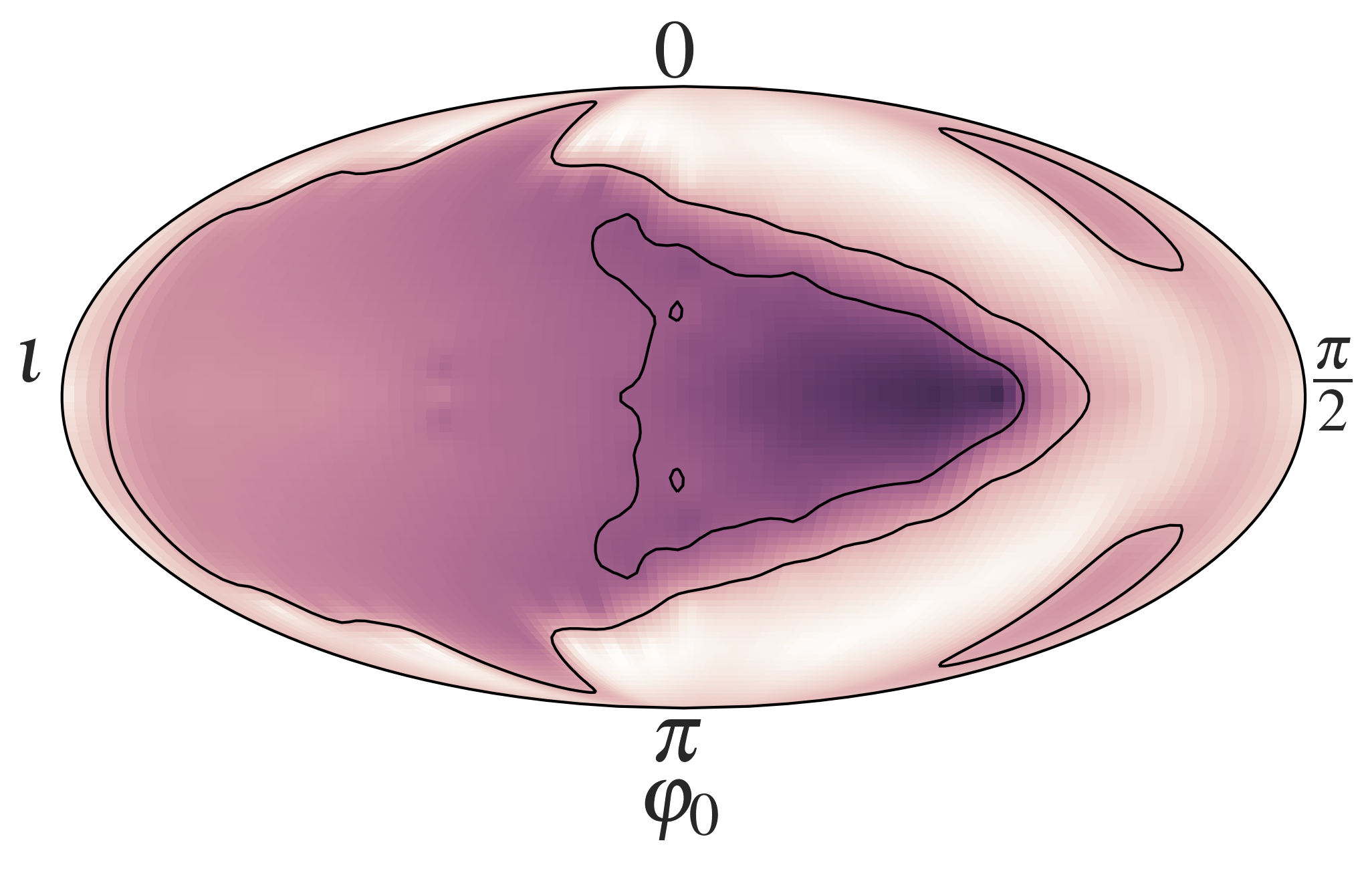}}
    {\includegraphics[scale=0.6]{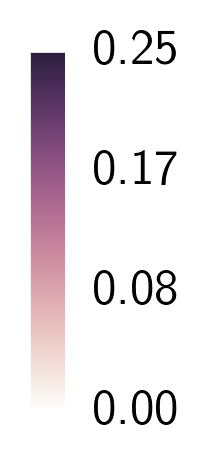}}
    
    \subfloat[]
    {\includegraphics[scale=\skymapscale]{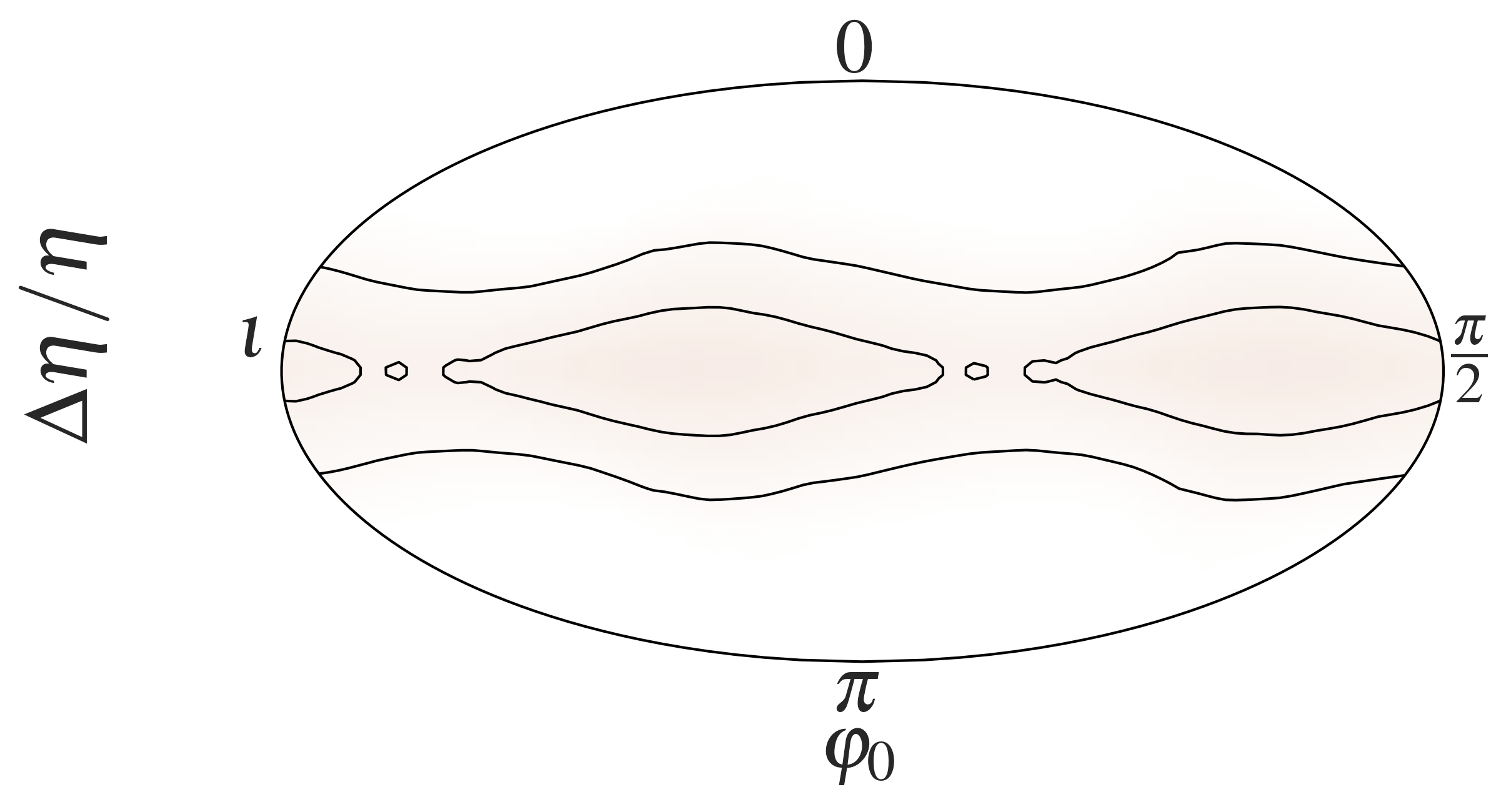}}
    \subfloat[]
    {\includegraphics[scale=\skymapscale]{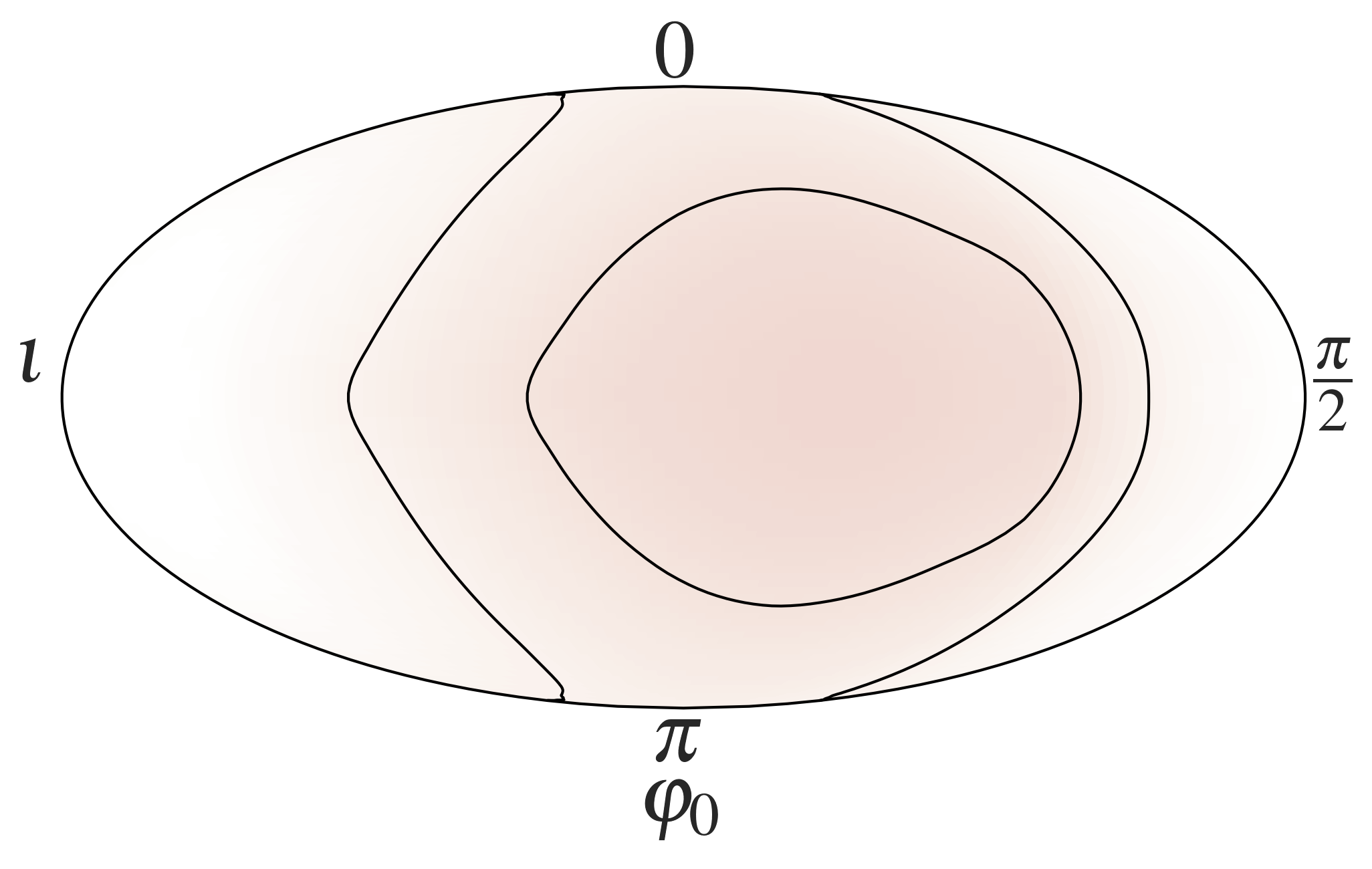}}
    \subfloat[]
    {\includegraphics[scale=\skymapscale]{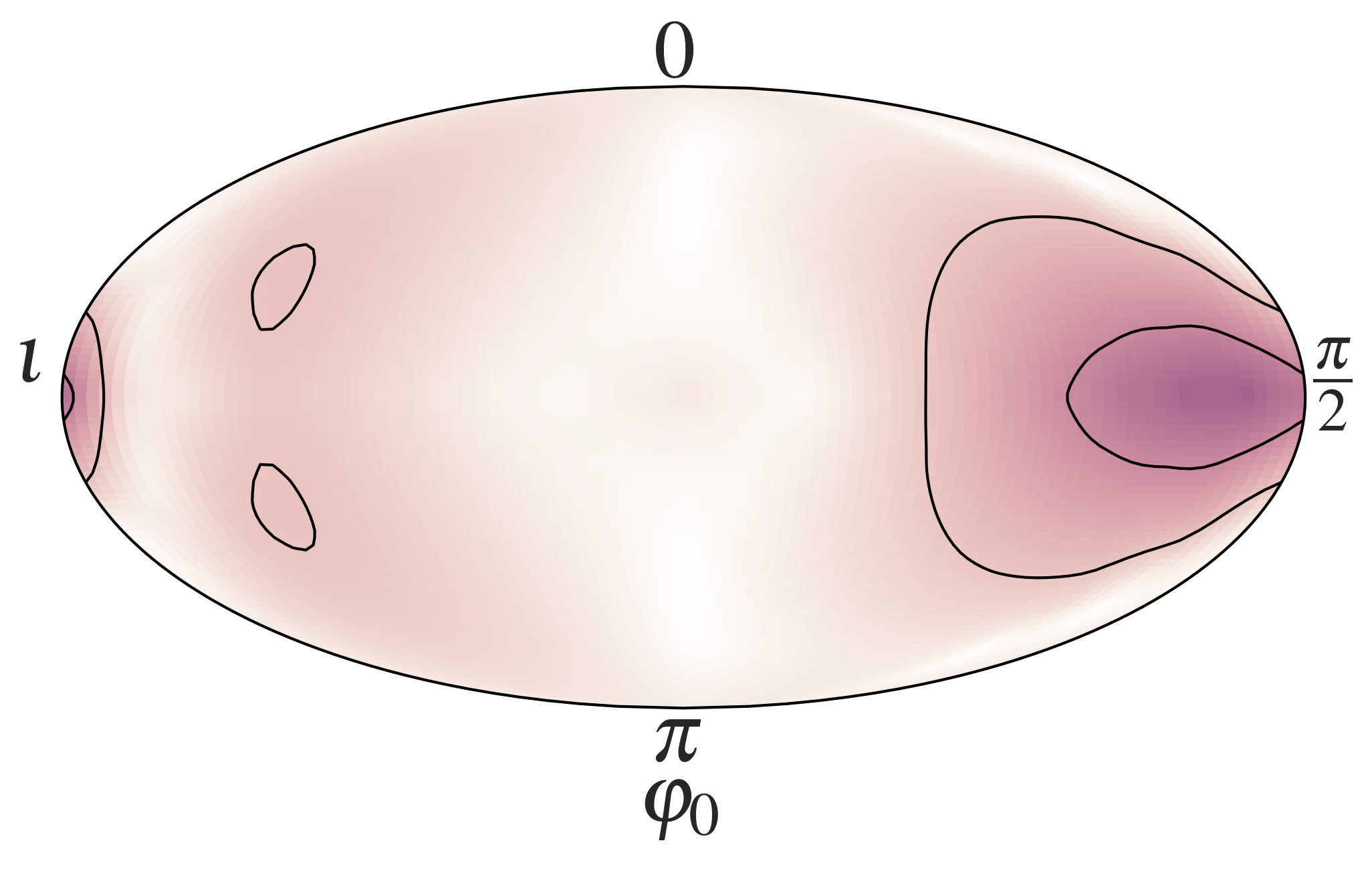}}
    \subfloat[]
    {\includegraphics[scale=\skymapscale]{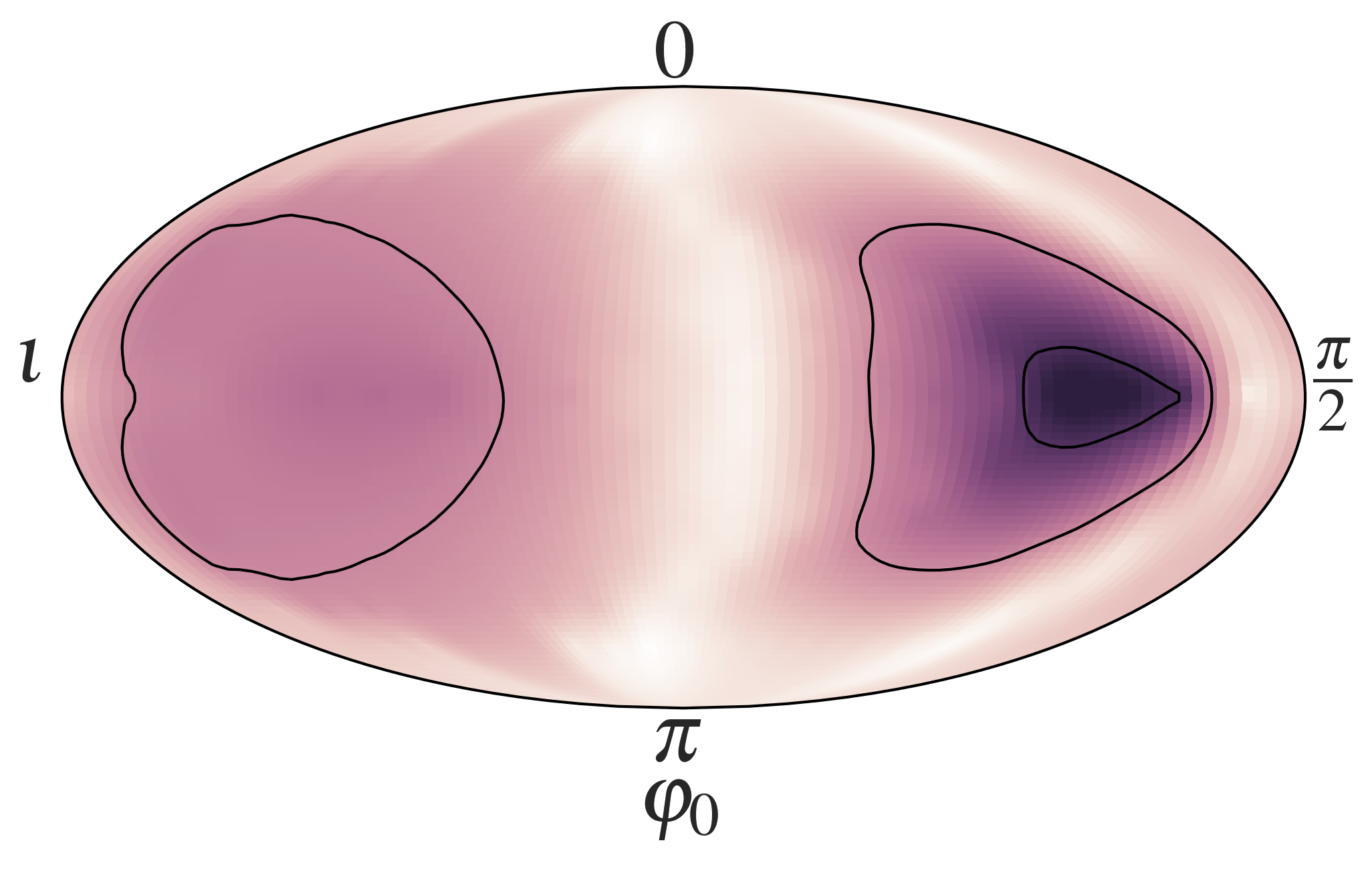}}
    {\includegraphics[scale=0.6]{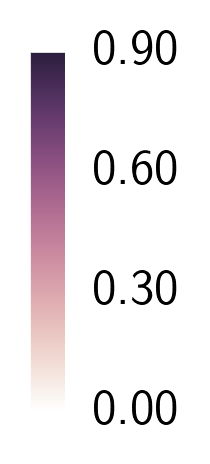}}

    \subfloat[$~q=1$, $\chi_{1z}=0.0$]
    {\includegraphics[scale=\skymapscale]{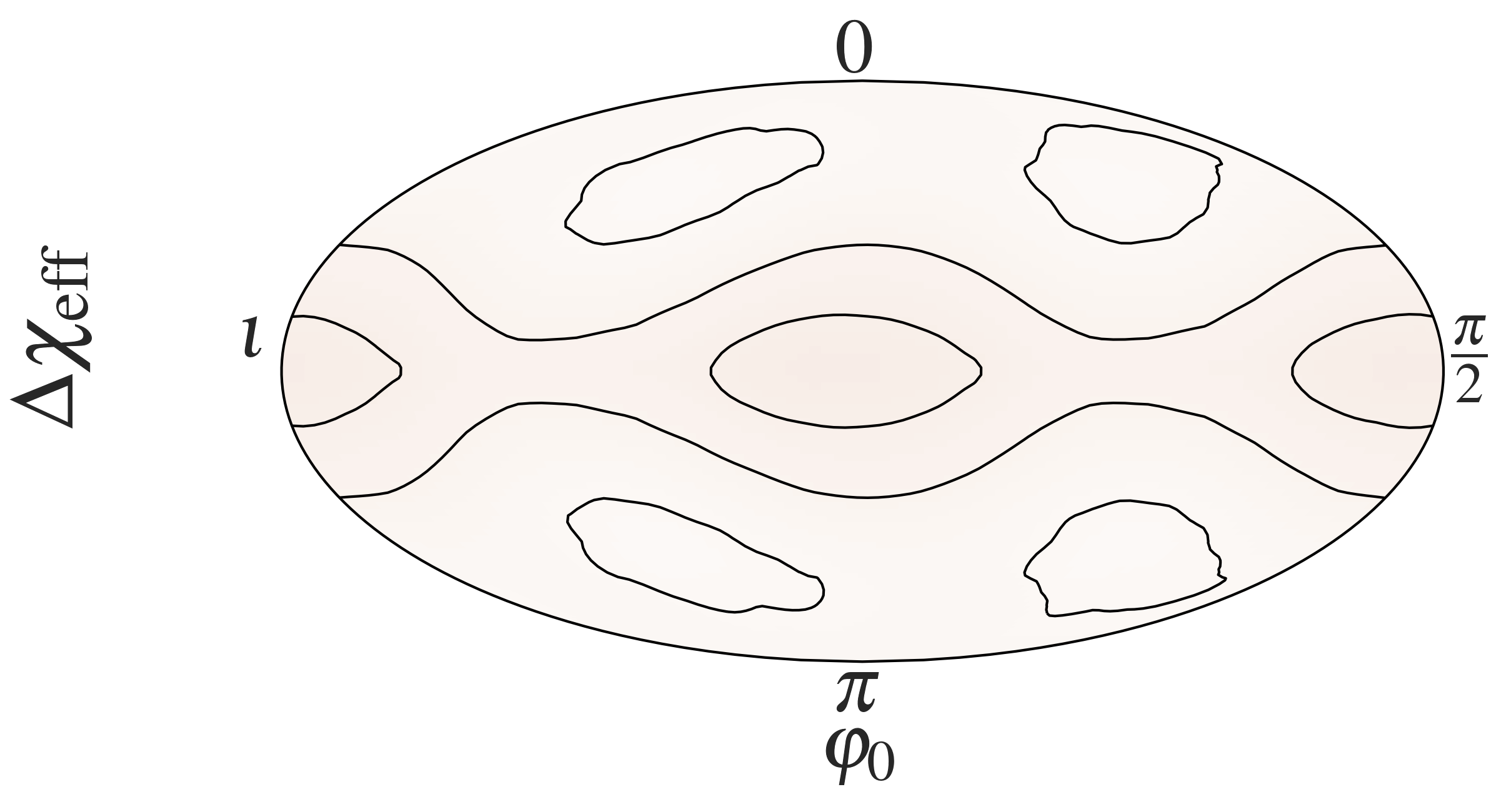}}
    \subfloat[$~q=8$, $\chi_{1z}=0.5$]
    {\includegraphics[scale=\skymapscale]{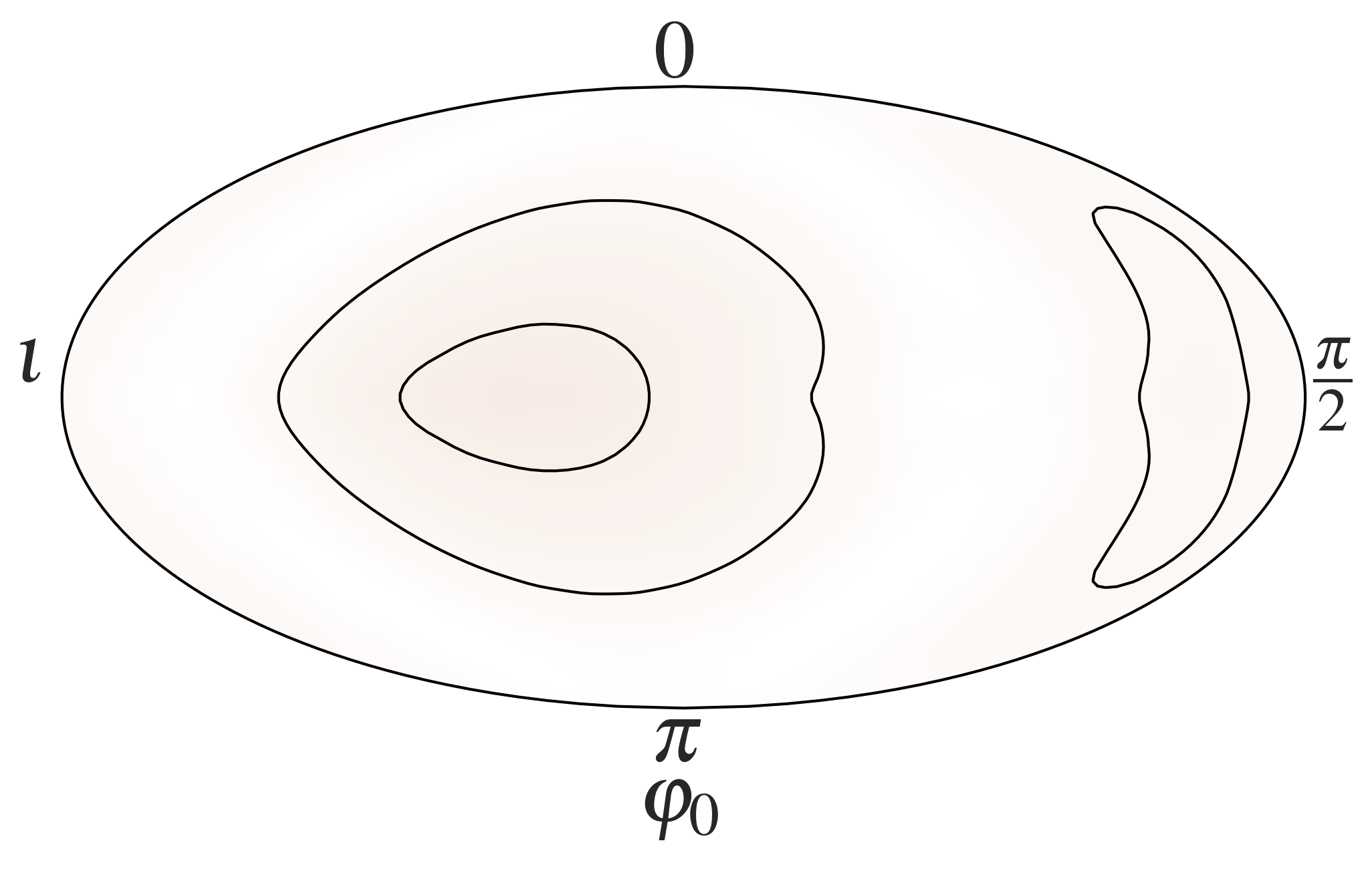}}
    \subfloat[$~q=8$, $\chi_{1z}=0.0$]
    {\includegraphics[scale=\skymapscale]{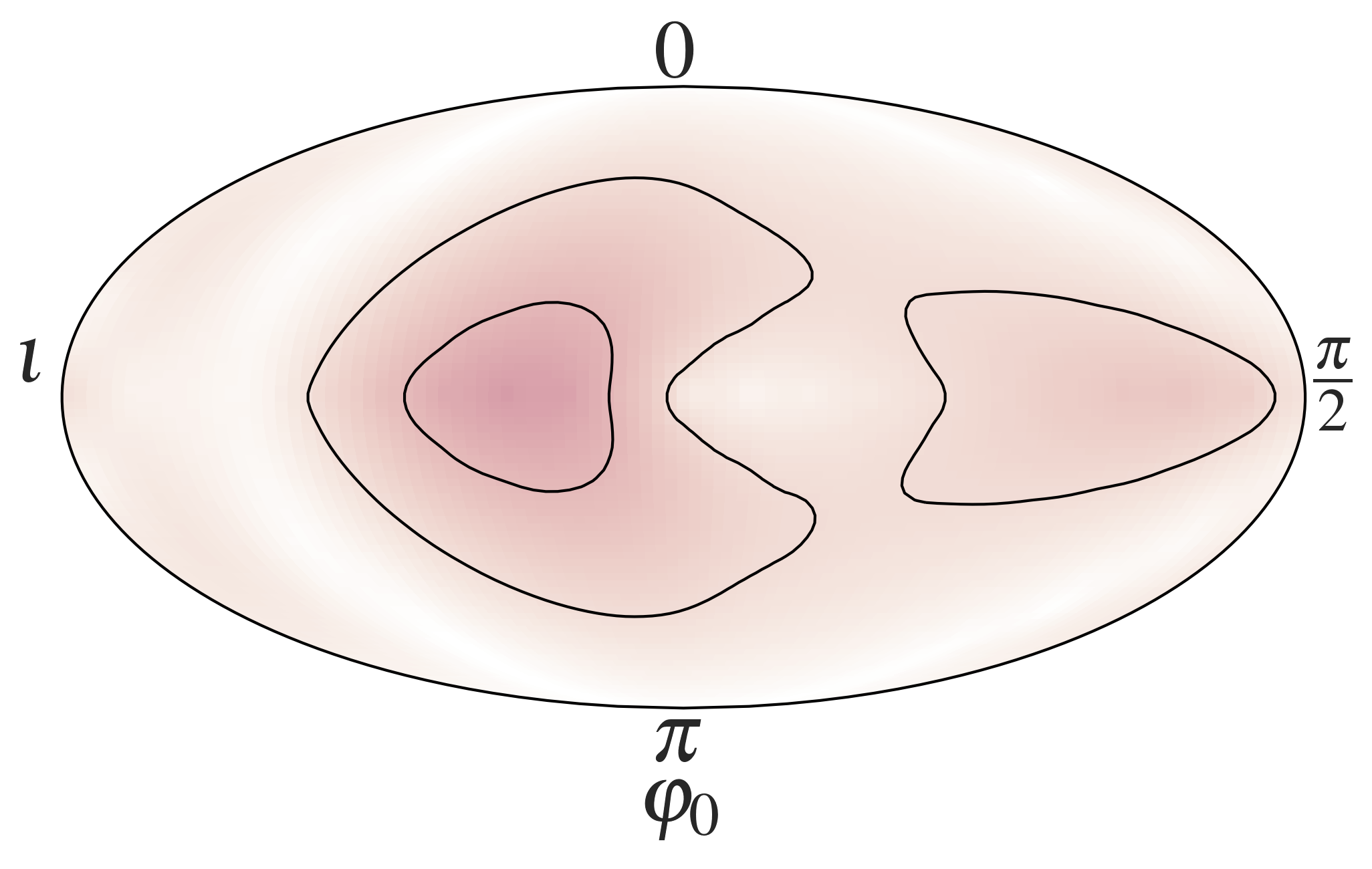}}
    \subfloat[$~q=8$, $\chi_{1z}=-0.5$]
    {\includegraphics[scale=\skymapscale]{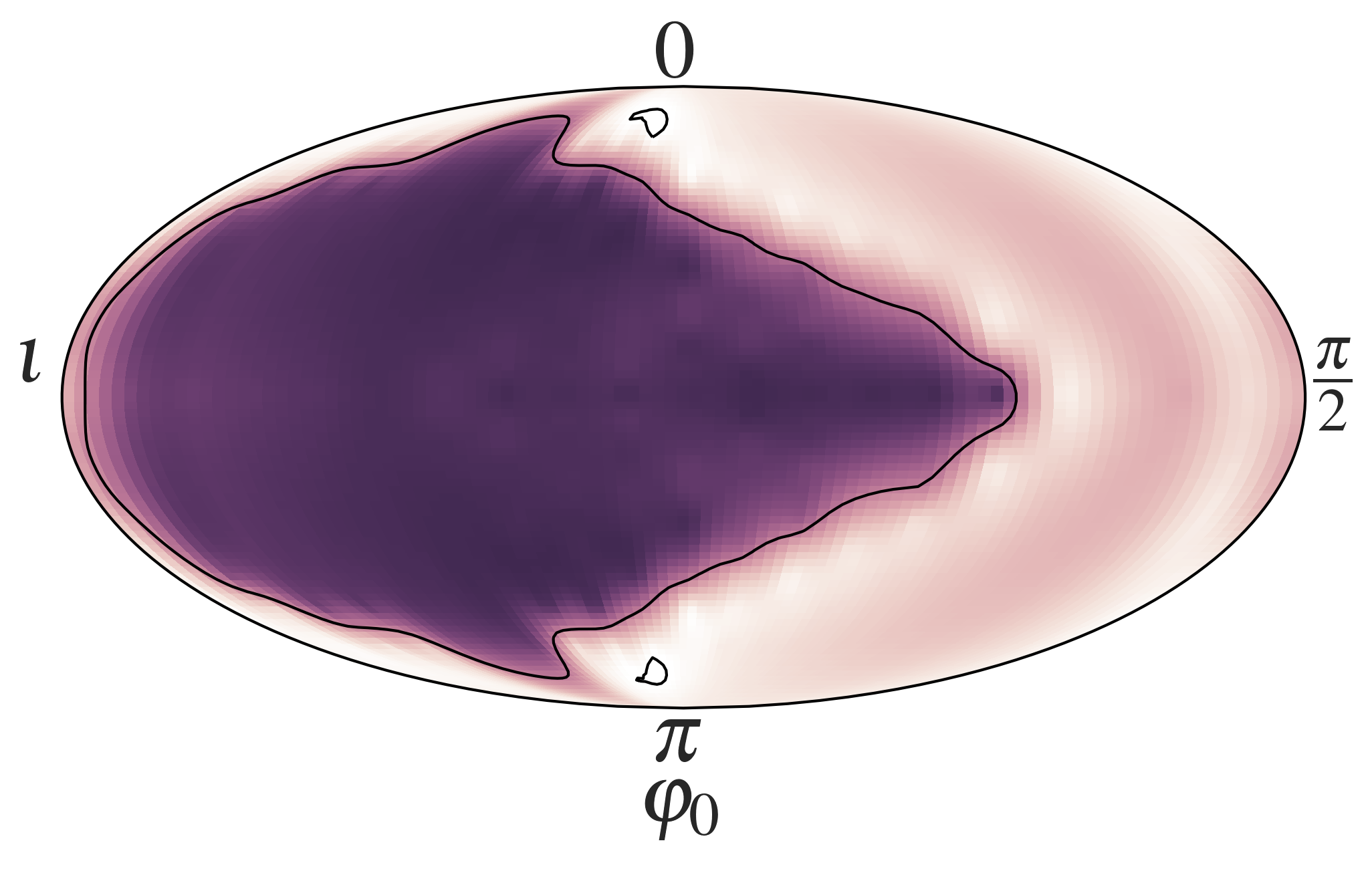}}
    {\includegraphics[scale=0.6]{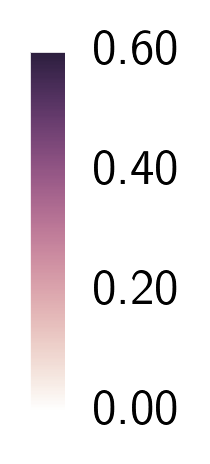}}
\caption{Systematic bias in the estimation of total mass $M$ (top panel), symmetric mass ratio $\eta$ (middle panel), and effective spin $\chieff$ (bottom panel), averaged over polarization angle $\psi$  for binaries with total mass $M=100~\Mo$. For $M$ and $\eta$, relative biases are shown, while for $\chieff$ absolute biases are shown. The y-axis shows the inclination angle $\iota$ in radians and the x-axis shows the initial phase of the binary $\varphi_0$ in radians. Different columns correspond to different mass ratios and spins of the larger black hole (the spin on the smaller black hole is 0 in all three cases).}
\label{fig:skymap_bias}
\end{center}
\end{figure*}

\begin{figure*}[tbh]
\begin{center}
\includegraphics[scale=0.45]{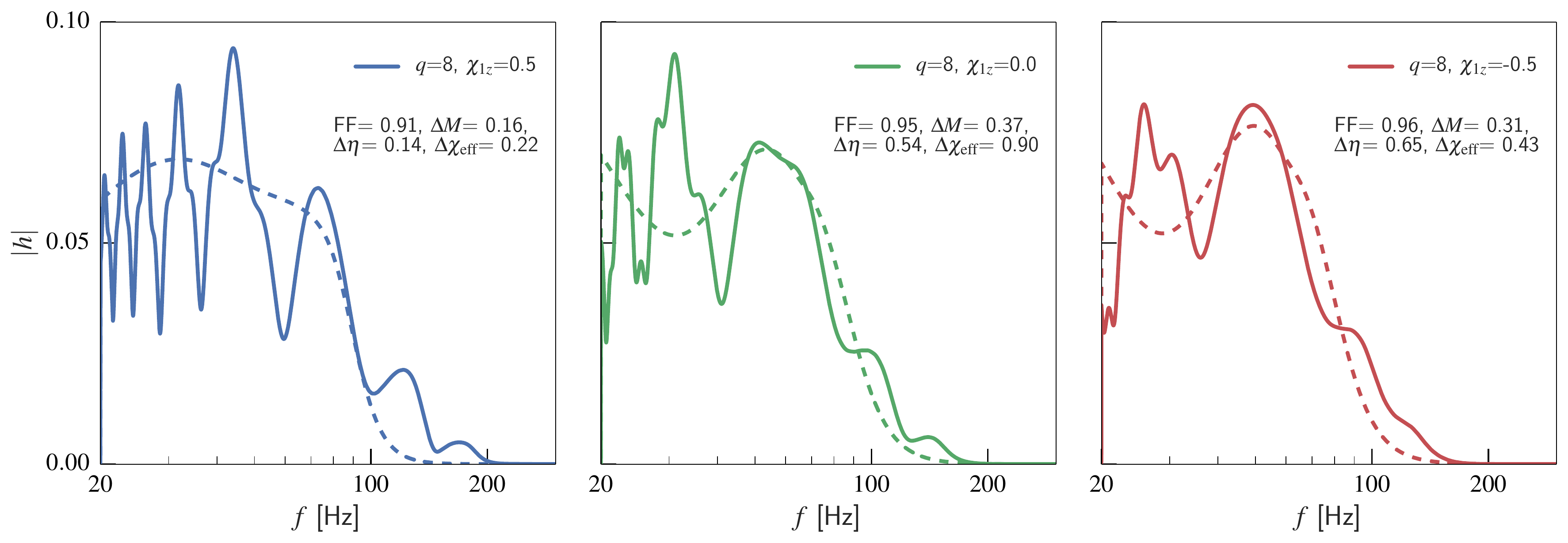}
\caption{Comparison of the frequency domain amplitudes of the ``full'' hybrid waveform containing 
subdominant modes (solid lines) and the best-match template waveforms containing only the 
quadrupole modes (dashed lines). The waveforms have been ``whitened'' according to the PSD 
used for match calculation and normalized such that the match with itself is unity. 
The orientation angles are chosen to be $\iota=\pi/4$, $\varphi_0=\pi$, $\psi=\pi/3$. 
The total mass is $M=200~\Mo$, and the mass ratio is $q = 8$. The legends show the spin of the 
larger black hole. The spin on the smaller black hole is zero in all three cases. 
The inset text shows the fitting factor, fractional biases in parameters $M$ and $\eta$ and 
absolute bias in parameter $\chieff$, at the best-match point. Particularly in the case of 
negative spin, where the observed signal is dominated by the ringdown, we see that the 
template is able to mimic the target, producing a reasonably good fitting factor. But this 
comes at the expense of larger parameter biases.}
\label{fig:best_match}
\end{center}
\end{figure*}
%

\section{Results and discussion}
\label{sec:results}

In this section, we evaluate the performance of the quadrupole-mode inspiral-merger-ringdown 
template family \TemplateName, against the ``full'' hybrid waveforms by computing the fitting 
factor of the template and inferring the parameter biases from the best-matched parameters. 
Figure~\ref{fig:skymap_ff_snr} shows the optimal SNR of the hybrid waveforms and fitting factor 
of the quadrupole-mode templates at different values of $\iota$ and $\varphi_0$ 
(averaged over the polarization angle $\psi$). Figure~\ref{fig:skymap_bias} shows the systematic 
bias in estimating parameters total mass $M$, symmetric mass ratio $\eta$, and effective 
spin $\chieff$, using the quadrupole-mode template family. It is clear that for the $q=1$ 
case (left column) the fitting factor is close to 1, and the systematic errors are negligible 
for all orientations, indicating the weak contribution of subdominant modes. For mass ratio 8, 
the fitting factor can be as low as $\sim 0.84$ for binaries that are highly 
inclined ($\iota \simeq \pi/2$) with the detector, where the contribution from nonquadrupole 
modes is the highest. However, these are the orientations where the SNR is the minimum 
(see Fig.~\ref{fig:skymap_ff_snr}). Similarly, the systematic biases are typically the 
largest (smallest) for the edge-on (face-on) configurations where the SNR is the 
smallest (largest). Hence GW observations are intrinsically biased toward orientations 
where the effect of nonquadrupole modes is minimum. This effect, in general, reduces the 
importance of nonquadrupole modes for a population of binaries that are oriented 
isotropically~\cite{Pekowsky:2013hm,Brown:2013hm,Capano:2013hm,Varma:2014hm}~\footnote{Note 
that this is an artifact of the limited horizon distance of the second-generation GW 
detectors. For the case of third generation GW detectors, binaries with practically all 
orientations will be detected, thus eliminating this selection bias; 
see, e.g.,~\cite{Vitale:2016aso}}.

Figure~\ref{fig:effparams} shows the \emph{ineffectualness} ($1-\FFe$) and effective biases in estimated parameters as a function of the total mass of the binary for different mass ratios and spins. For total mass $M$ and symmetric mass ratio $\eta$, fractional biases are shown while for $\chieff$ absolute biases are shown\footnote{In the case of anti-symmetric spin parameter $\mchieff$, the biases are dominated by the bias in the quadrupole mode itself. This is expected as previous studies have shown that LIGO can only estimate $\chieff$ to a good accuracy. Therefore we do not consider biases in $\mchieff$ in this study.}. Solid (dashed) lines correspond to the case where ``full'' (quadrupole-only) hybrid waveforms are used as target waveforms. The template family in both cases contains only the quadrupole mode. The difference between the solid and dashed lines indicates the effect of ignoring sub-dominant modes for detection and parameter estimation. Note that many of the dashed lines lie below the scale of these plots and are not displayed.

Previous studies~\cite{Pekowsky:2013hm,Brown:2013hm,Capano:2013hm,Varma:2014hm,CalderonBustillo:2016hm} have shown that the effects of subdominant modes become important for binaries with high masses and large  mass ratios. At large mass ratios, subdominant modes are excited by a larger extent due to higher asymmetry.  For high masses, the observed signal is dominated by the merger, during which sub-dominant modes are excited prominently. Consistent with our expectation, in Fig.~\ref{fig:effparams}, the solid lines show that, in general, the ineffectualness and effective biases increase with increasing mass ratio and with increasing mass. We also see a clear separation of the solid and dashed lines for large mass ratios and high masses, illustrating the effect of neglecting nonquadrupole modes. 

Figure~\ref{fig:effparams} also reveals an interesting dependence of the effect of 
nonquadrupole modes on the spins. For binaries with aligned, zero, and antialigned spins, 
the ineffectualness peaks at total masses of $M \sim 300 M_\odot$, $M \sim 150 M_\odot$, 
$M \sim 100 M_\odot$, respectively~\footnote{Note that this is not true for the $q \simeq 1$ 
cases. For these, since the mismatches are quite small $\sim 10^{-3}$, several competing effects 
are playing out.}. This is roughly the mass range where the observed signal is dominated by 
the late inspiral and merger -- the phase where the higher modes are excited most prominently. 
For binaries with antialigned spins, merger happens at relatively lower frequencies, while, 
for the case of aligned spins, merger happens at relatively higher frequencies, owing to 
the ``orbital hangup''~\cite{Campanelli_2006orbitalhangup, Hannam_2008orbitalhangup} effect. 
Since frequencies are scaled inversely to the total mass of the system, this creates the 
mass dependence of the ineffectualness that we describe above. For very high masses, 
the observed signal will contain only the ringdown phase. Due to the smaller bandwidth and 
the relatively simpler structure of the ringdown signal, the quadrupole-only templates are 
likely to be able to mimic the full ringdown signal relatively well, at the cost of 
considerable systematic errors (see Fig.~\ref{fig:best_match} for an example). Hence, we 
anticipate the effectualness of the quadrupole-mode templates to go up at very high masses. 
This effect should start dominating the effectualness patterns at relatively lower masses 
for binaries with antialigned spins. Consistent with our expectation, we see in 
Fig.~\ref{fig:effparams} (top left panel) that for a given mass ratio, at low masses, 
binaries with negative spins have higher ineffectualness but as the mass increases there 
is a crossover point beyond which binaries with positive spins have higher ineffectualness. 
While for positive spins, the ineffectualness continues to increase with total mass, 
for zero spins the ineffectualness plateaus and for negative spins it reaches a maximum 
value and starts deceasing beyond that point. We see from Fig.~\ref{fig:effparams} that 
this trend of larger (smaller) effectualness for negative (positive) spins at high 
masses ($M \gtrsim 100 M_\odot$) is achieved at the cost of larger (smaller) systematic 
biases in the estimated parameters.

\begin{figure*}[htb]
\centering
    \hspace*{1.0cm}{\includegraphics[scale=\minsnrscale]{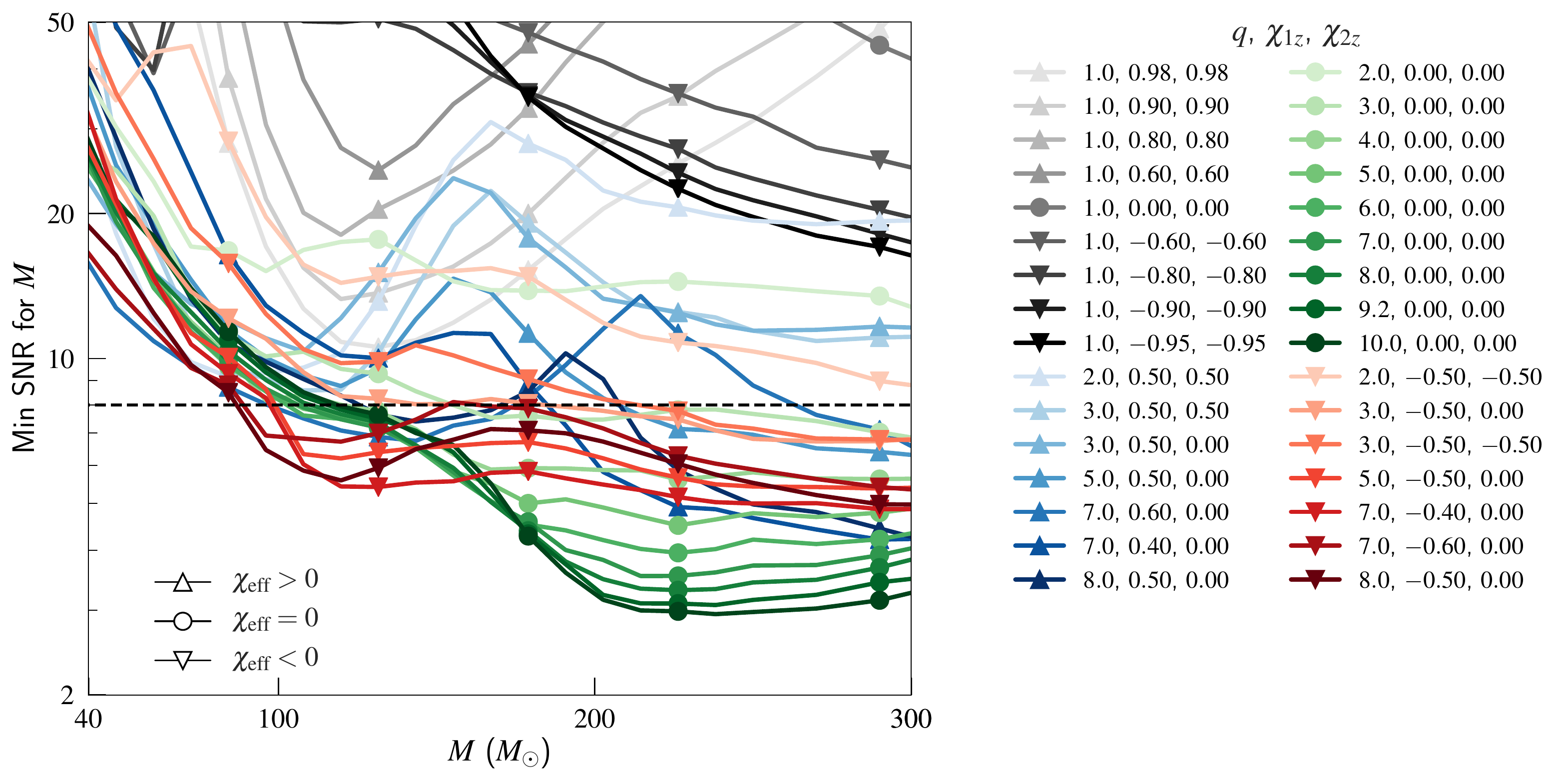}} \\
    {\includegraphics[scale=\minsnrscale]{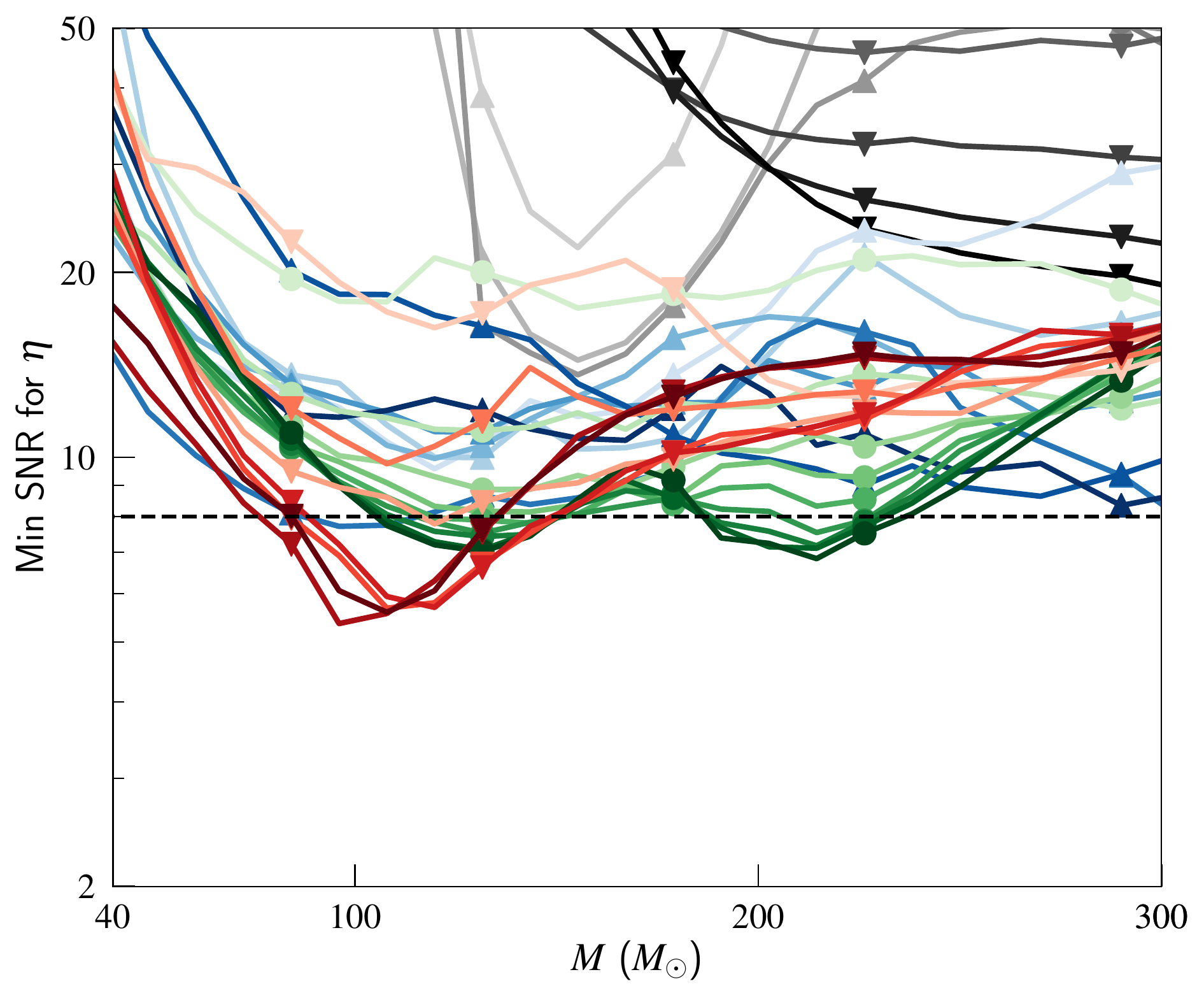}}
    {\includegraphics[scale=\minsnrscale]{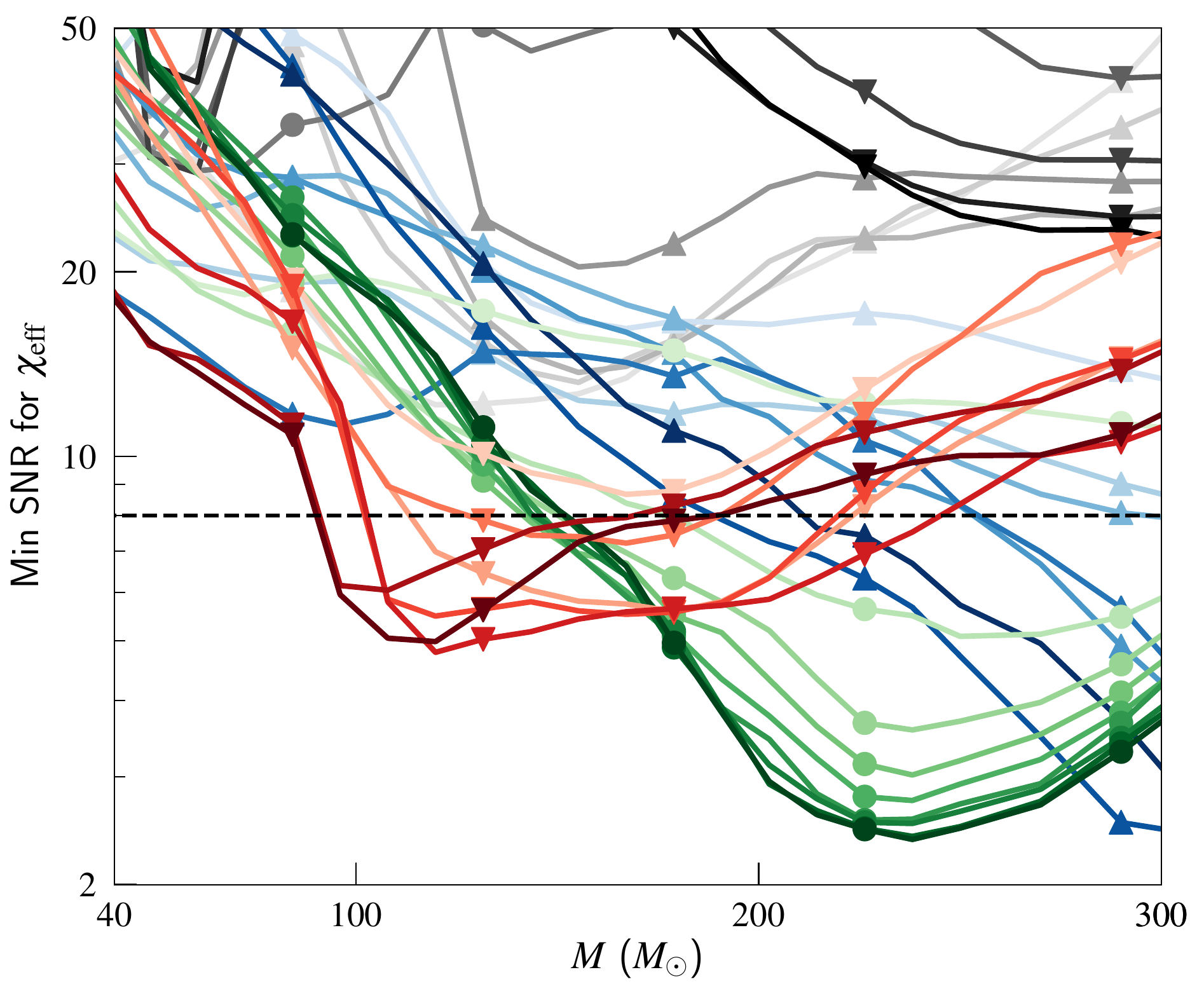}}
\caption{Lowest SNR (orientation-averaged) at which the statistical errors are low 
enough to equal the effective systematic bias in parameters $M$, $\eta$, and $\chieff$, 
when using quadrupole mode templates to estimate the parameters of hybrid waveforms including 
all modes. A dashed black line is used to denote minimum orientation-averaged SNR of 
$8$ (optimal orientation SNR of $20$).}
\label{fig:minsnr}
\end{figure*}

We set $\FFe \geq 0.965$ (which corresponds to a $\sim10\%$ loss in detection volume for a 
fixed SNR threshold) as the benchmark for the relative importance of nonquadrupole 
modes in detection. This is shown by the dashed black line in the top-left panel of 
Fig.~\ref{fig:effparams}. Figure~\ref{subfig:summary_det} summarizes the region in 
the parameter space where the loss of detectable volume (at a fixed SNR threshold) 
due to neglecting nonquadrupole modes is greater than 10\%. For the case of negative spins, 
even at large mass ratios, we see that subdominant modes are important for detection only 
over a range of masses ($M \sim 75 - 150 M_\odot$). For binaries with positive and zero 
spins, we anticipate that the upper limit of total mass where the higher modes are important 
is above $300 M_\odot$, the highest mass that we consider in this study. Based on 
Fig.~\ref{subfig:summary_det}, we expect the quadrupole mode templates to be fully effectual 
for detection either when $q \lesssim 4$ or when $M \lesssim 70 M_\odot$ (irrespective 
of spins), considering a population of binaries distributed with isotropic orientations. 
We note that the region in which subdominant modes become important for detection is the 
smallest (largest) for negative (positive) spins. 

Figure~\ref{subfig:summary_det} also shows the region in the parameter space (marked by the green dashed line) where subdominant modes are important for the detection of nonspinning binaries when nonspinning quadrupole mode templates are used, obtained in our previous study~\cite{Varma:2014hm}. We see that the use of quadrupole  mode templates with nonprecessing spins has helped us to reduce the region in the parameter space where subdominant modes cause unacceptable loss in the detection volume. This is consistent with our expectation, as two additional parameters (spins) in the templates allow them to achieve higher fitting factors with the target signals, at the cost of a larger bias in the best-matched template parameters.

In order to gauge the relative importance of the systematic errors shown in Fig.~\ref{fig:effparams}, we compare them against the expected statistical errors from the quadrupole-mode template family \TemplateName\ (computed using Fisher matrix formalism). Figure~\ref{fig:minsnr} shows the minimum SNR (orientation-averaged) at which the $1\,\sigma$ statistical errors become low enough to equal the systematic errors. (Note that statistical errors are inversely proportional to the SNR.) We see that, at high masses, the systematic errors start to dominate the error budget for orientation-averaged SNRs as low as $3$. In this study, whenever the systematic errors are less than the statistical error for an orientation-averaged SNR of 8 (horizontal black dashed line in Fig.~\ref{fig:minsnr}), we regard the quadrupole-mode templates to be faithful for parameter estimation \footnote{Note that, when full mode templates are employed in the parameter estimation, the statistical errors are expected to go down in general, due to the increased amount of information in the waveform~(see, e.g., \cite{VanDenBroeck:2006qu}). We do not consider this effect here.}. 

Figure~\ref{subfig:summary_param} summarizes the region in the parameter space where this minimum orientation-averaged SNR is less than or equal to $8$ for estimation of any of $M$, $\eta$ or $\chieff$. We exclude any cases where the systematic biases are dominated by the biases in the quadrupole mode itself. We note that the region in which subdominant modes become important for parameter estimation is smallest (largest) for positive (negative) spins. This trend is opposite to what we see in Fig.~\ref{subfig:summary_det} for detection. This is because, at high masses negative spin binaries have higher effectualness than positive spin binaries, which is achieved at the cost of higher systematic biases. We remind the reader that, for spins of higher magnitude than considered in this study (i.e. $|\chieff| > 0.5 $ for $q\geq2$), we expect the shaded regions in Fig.~\ref{fig:summary_fig} to expand or reduce depending on the spin; the contours that we draw are indicative demarcations only. For greater aligned spins, the shaded region for detection should expand and the shaded region for parameter estimation should reduce. The opposite trend is expected for greater antialigned spins. Figure~\ref{subfig:summary_param} also compares these results with the results obtained in our previous study~\cite{Varma:2014hm} (dashed green line) using nonspinning quadrupole-only templates against nonspinning ``full'' target waveforms. We see that the use of spinning templates essentially increases the region where the parameter estimation bias is dominated by systematic errors. 

\section{Conclusion}
\label{sec:conclusion}
We studied the effects of sub-dominant modes in the detection and parameter estimation of GWs from black hole binaries with nonprecessing spins using Advanced LIGO detectors. The effect of sub-dominant modes on detection is quantified in terms of the effective detection volume (fraction of the optimal detection volume that the suboptimal search is sensitive to, for a given SNR threshold) and the effect on parameter estimation in terms of the effective bias (weighted average of the systematic errors for different orientations) in the estimated parameters. We compared quadrupole-mode templates with target signals (hybrid waveforms constructed by matching NR simulations describing the late inspiral, merger and ringdown with PN/EOB waveforms describing the early inspiral). These signals contained contributions from all the spherical harmonic modes up to $\ell = 4$ and $-\ell \leq m \leq  \ell$ except the $m=0$ modes. 

Our study considered black hole binaries with total masses $40\Mo \leq M \leq 300\Mo$, 
mass ratios $1 \leq q \leq 10$, and various spins including  
$\chieff \sim -0.5, 0, 0.5$ ($|\chieff| \leq 0.98$ for $q=1$). The results are appropriately 
averaged over all angles describing the orientation of the binary (the results are not 
explicitly averaged over the sky location because both the fitting factors and systematic 
biases are only weakly dependent on the sky location \footref{Ftnote:angles}). 
Figure~\ref{fig:summary_fig} shows the regions in the parameter space where the contribution 
from nonquadrupole modes is important for GW detection and parameter estimation. 
In general, neglecting subdominant modes can cause unacceptable loss of SNR and unacceptably 
large systematic errors for binaries with high masses and large mass ratios. For a given mass 
ratio, subdominant modes are more important for positive (negative) spins for detection 
(parameter estimation). As compared to our previous study restricted to the case of 
nonspinning binaries, we see that the use of quadrupole mode templates with nonprecessing 
spins, enhances the effectualness for detection, but extends the region where systematic 
errors dominate. 

Note that the scope of our study was rather restricted -- while we conclude that subdominant 
mode templates are likely to improve the detection rates of binary black holes in certain 
regions in the parameter space (high mass and large mass ratios), a proper characterization 
of this will require characterizing the associated increase in the false alarm rate also 
(see, e.g., Ref.~\cite{Capano:2013hm}). Also, we did not study the effect of neglecting 
nonquadrupole modes on signal-based vetoes such as the ``chi-square'' veto~\cite{Allen:2004gu}. 
Similarly, we have only investigated the region in the parameter space where the use 
of the quadrupole-only template would introduce systematic errors that are larger than 
the expected statistical errors. However, the use of full-mode templates in parameter 
estimation is likely to reduce the statistical errors, owing to the increased information 
content in the waveform. We have not explored this aspect of the problem here. The expected 
statistical errors were estimated using the Fisher matrix formalism. Since these error bounds 
are lower limits, our estimates on the region of the parameter space where the systematic errors 
are negligible should be treated as conservative estimates. 
We conclude that subdominant modes are important for parameter estimation when the systematic 
errors are greater than $1\sigma$ statistical errors at a sky and orientation averaged SNR of 8. 
If more stringent criteria are applied, our shaded regions in Fig.~\ref{subfig:summary_param} 
would widen. Also, note that we restricted our study to the case of binaries with 
nonprecessing spins. Astrophysical black hole binaries may have generic spin orientations. 
It is not clear how our conclusions hold in the case of precessing spins 
(see Ref.~\cite{Bustillo2016_hm} for some recent work in this direction). We leave some of 
these investigations as future work. 

\appendix 
\section{Comparison with Bayesian parameter estimation}
\label{sec:comparison_bayesian}

In this paper, we tried to quantify the loss of detection efficiency due to neglecting 
subdominant modes by computing the fitting factors of the dominant-mode templates with target 
signals including the effect of subdominant modes. Systematic errors in parameter estimation were 
computed by comparing the parameters of the ``best-matched'' subdominant-mode templates with 
the true parameters of the target signals, while statistical errors are computed from the 
Fisher information matrix. Since these calculations are computationally inexpensive, this 
allows us to study the impact of subdominant modes over the entire parameter space of interest, 
after averaging over extrinsic parameters such as the orientation angles. However, we know 
that the inverse of the Fisher matrix provides a \emph{lower bound} of the statistical errors 
in the parameter estimation~\cite{cramer1999mathematical,MR0015748}. In order to verify that 
our simplified estimates of the statistical and systematic errors give a good approximation 
to the true errors, we compare our estimates of the systematic and statistical errors with 
those derived from full Bayesian parameter estimation for one sample case. 

\setlength{\tabcolsep}{1.3em}
\begin{table*}[hbt]
 \begin{tabular}{c  c  c  c  c  c  c} 
\toprule
 Method & $M_\mathrm{bias}$ & $\eta_\mathrm{bias}$ & $\chi^{\mathrm{eff}}_\mathrm{bias}$ 
 & $\delta M$ & $\delta \eta$ & $\delta \chieff$\\ [0.5ex] 
\midrule
 Bayesian &  $7.6 \times 10^{-1} \Mo$ & $2.2 \times 10^{-3}$ & $4.3 \times 10^{-2}$ 
 & $2.8 \Mo$ &  $2.4 \times 10^{-3}$ & $1.3 \times 10^{-1}$\\ 
 FF/Fisher &  $2.6 \times 10^{-1} \Mo$ & $2.7 \times 10^{-3}$ & $5.4 \times 10^{-2}$ 
 & $1.9 \Mo$ &  $9.5 \times 10^{-3}$ & $7.0 \times 10^{-2}$\\ 
 \hline
\end{tabular}
\caption{
Comparison of systematic and statistical biases as predicted by a full Bayesian parameter 
estimation (top row) study with a fitting factor/Fisher matrix study that is used in this paper (bottom row). The first three columns show 
the absolute systematic biases and the next three columns show the statistical errors  
in the estimation of total mass $M$, symmetric mass ratio $\eta$ and effective spin $\chieff$. 
For the Bayesian study (top row), systematic biases are inferred 
from the peaks of the posterior distributions and the statistical biases are given by 
	the widths of $68\%$ credible intervals. In the bottom row, the systematic biases are inferred from the best-match 
parameters and the statistical biases are given by $1\sigma$ errors from a Fisher matrix 
study. 
}
\label{Table:PEvsFisher}
\end{table*}

We create a simulated data stream by injecting a numerical-relativity waveform from the 
SXS waveform catalog~\cite{Mroue:2013xna, SXS-Catalog, Schmidt:2017btt} into colored Gaussian 
nose with the power spectrum of Advanced LIGO. The injected waveform (SXS:BBH:0307) has 
the mass ratio $m_1/m_2=1.228$, aligned spins $\chi_{1} = 0.32$, $\chi_{2} = -0.5798$, and has 
a SNR of $\sim 25$. We estimate the posterior distributions of of the masses and spins using 
the \textsc{LALInferenceNest} code~\cite{Veitch:2009hd, Veitch:2014wba} that is part of the 
LSC Algorithms Library~\cite{LAL}. We compare the maximum a posteriori probability (MAP) estimates 
with the true parameters, which provides us an estimate of the systematic bias. Similarly, 
the width of the 68\% credible regions provides us an estimate of the statistical errors. 
These estimates are compared with the same estimated using the methods that we use in 
the paper. Table~\ref{Table:PEvsFisher} provides a comparison between these independent 
estimates. We see that, for the parameters that we consider, the two different estimates 
are in reasonable agreement. Although this provides some confidence in our results, extensive 
comparisons with Bayesian estimates over the full parameters space are required to confidently 
establish the accuracy of our approximate results. We leave this as future work.

\smallskip \acknowledgments 
We are indebted to the SXS Collaboration for making a public catalog of numerical-relativity 
waveforms, and to Chandra Kant Mishra for sharing a notebook of post-Newtonian waveforms. 
We thank Abhirup Ghosh, Chandra Kant Mishra, Sascha Husa, Mark Hannam, Michael P\"urrer, and 
Patricia Schmidt for useful discussions. We also thank Richard O'Shaughnessy, 
B. S. Sathyaprakash, Prayush Kumar, and the anonymous referee for several useful comments on 
the manuscript. P.~A.'s research was supported by the AIRBUS Group Corporate Foundation through 
a chair in ``Mathematics of Complex Systems'' at the International Centre for Theoretical 
Sciences (ICTS); by a Ramanujan Fellowship from the Science and Engineering Research 
Board (SERB), India; by the SERB FastTrack fellowship SR/FTP/PS-191/2012; by Indo-US 
Centre for the Exploration of Extreme Gravity funded by the Indo-US Science and Technology 
Forum; and by the Max Planck Society and the Department of Science and Technology, India, 
through a Max Planck Partner Group at ICTS. V.~V.'s research was supported by NSF 
Grant No. PHY-1404569 to Caltech and the Sherman Fairchild Foundation. Computations were 
performed at the ICTS clusters Mowgli, Dogmatix, and Alice.

\bibliography{HigherModesWithSpins}
\end{document}